# Low Temperature Two Fluid State in SmB$_6$

Sayantan Ghosh[1], Sugata Paul[1], Tamoghna Chattoraj[1], Ritesh Kumar[1], Zachary Fisk[2], S. S. Banerjee[1,†]

[1]*Indian Institute of Technology Kanpur, Kanpur, Uttar Pradesh 208016, India.*

[2]*Department of Physics and Astronomy, University of California at Irvine, Irvine, CA 92697, USA.*

Corresponding author email : [†]satyajit@iitk.ac.in

**Abstract :** Comprehensive study using DC transport, specific heat, magnetization, and two-coil mutual inductance measurements unveils an understanding of three temperature regimes in SmB$_6$: (i) $T \geq T^*$ (~66 K), (ii) $T_g$ (~40 K) $\leq T < T^*$, and (iii) $T < T_g$. Onset of Kondo breakdown below $T^*$ releases disorder-driven magnetic fluctuations, which splits the bulk (~116 K) and surface Kondo temperature ($T_k^s \approx 7$ K). Below $T_g$, as magnetic fluctuations subside, surface Kondo screening revives, stabilizing the topological surface state and generating an in-gap feature (~2.2 meV) across which Dirac-like carriers are excited. Nyquist impedance analysis reveals a crossover from purely capacitive to capacitive-inductive behavior, signalling a disorder-driven two-fluid phase of heavy quasiparticles and light, high-mobility carriers below $T_g$. We identify a characteristic length scale, $\mathcal{L}_{\nu_0}(T)$, associated with the high-mobility phase, exhibiting an almost divergent trend below $T_k^s$. These findings underscore the complex nature of surface conducting state in SmB$_6$.

**Introduction:**

Strong electronic correlations and band topology are seen as intertwined factors governing exotic quantum phases in condensed matter physics. Phenomena in materials, characterized by narrow or flattened electronic bands where correlation effects are significantly enhanced, can be effectively captured by frameworks rooted in Kondo physics [1]. This interplay, in which electronic band topology and strong electron correlation effects together drive novel quantum states, paves the way toward a unified understanding of diverse quantum materials. Samarium hexaboride ($SmB_6$) is among the rare materials that simultaneously exhibit strong correlation-driven Kondo physics as well as topological insulating characteristics. Its features have challenged our understanding, leading to debates and unresolved issues which promise to expand the boundaries of condensed matter physics. $SmB_6$ has emerged as a prototypical Kondo insulator (KI) and a candidate for a topological Kondo insulator (TKI) [2-10]. Here Kondo hybridization between itinerant $5d$ and localized $4f$ electrons generates a narrow bulk Kondo gap (10–20 meV), marking the onset of the Kondo insulating phase [2,4]. The band inversion arising from the hybridization of the odd-parity $f$-band and even-parity $d$-band, along with strong spin-orbit coupling, imparts nontrivial $\mathbb{Z}_2$ topological insulator characteristics to its band structure [10,11]. Bulk-boundary correspondence predicts surface states featuring topological Dirac-like dispersion within the Kondo gap. The resistivity ($\rho$) rapidly increases below $\sim 30 \pm 15$ K [5,12-16] and plateaus out below 4 K. Transport [12-14], quantum oscillations [17-20] and spectroscopic studies [15,16,21-26] in 4 K regime suggest 2D in-gap states. However, the existence of pure 2D topological surface conducting states remains debatable [4]. The absence of SdH oscillations and variations in dHvA oscillation frequencies have led to conflicting interpretations: either a 2D metallic Fermi surface indicating surface conduction or a bulk 3D Fermi surface with charge-neutral fermions or, $SmB_6$ hosts both bulk and surface excitations [18-20,27-31]. Additionally, the effective mass of surface states - a crucial metric for correlation physics - exhibits $\sim 0.2 - 1000$ times free-electron mass variation across different measurements [4]. Disorder effects in $SmB_6$ further complicate the picture by resulting in polar surfaces [32], weakening topological protection, disrupting spin-momentum locking, localizing surface states and even introducing non-topological in-gap states [2,33-35]. Despite these, surprisingly, low-temperature $\rho$ saturation is unaffected by sample quality [2]. These observations highlight the necessity for a deeper exploration of surface conduction properties in $SmB_6$.

The non-contact two-coil mutual inductance (TCMI) technique [36-38] effectively separates bulk and surface contributions, unlike electrical transport measurements, where they are intertwined. Consistent with prior studies [26], TCMI investigations of SmB$_6$ [38] identify three temperature regimes: (i) $T \geq T^*$ (~66 K), (ii) $T_g$ (~40 K) $\leq T < T^*$, and (iii) $T < T_g$. While DC electrical conductivity saturates below 4 K, AC conductivity ($\sigma(\nu)$) exhibits distinctive features for $T < T_g$, with $\sigma(\nu) \propto \nu^{0.5}$ [38]. In contrast, for Bi$_2$Se$_3$, where highly conducting topological surface states dominate conduction, $\sigma(\nu) \propto \nu^{(\sim 1)}$ [37-39]. Our study explores the different $T$ regimes in SmB$_6$, revealing disorder-driven deviations from ideal Kondo insulating behavior, with a split between bulk and surface Kondo temperatures. We identify an in-gap feature (~2.2 meV) linked to presence of Dirac point within the Kondo gap. A disorder-induced two-fluid conduction mechanism emerges at low temperatures from surface, involving light, high-mobility Dirac-like carriers and heavy quasiparticles. Additionally, we identify a characteristic length scale associated with the high-mobility fluid phase, exhibiting an almost diverging trend at low $T$. All these suggest a heterogenous two fluid phase with heavy and light quasiparticles contributing to surface conductivity of SmB$_6$.

**Results:**

This study investigates SmB$_6$ through DC transport, specific heat, DC magnetization, and TCMI techniques [37,38]. We studied two SmB$_6$ single-crystal samples: sample I (s$^I$, 0.9 × 0.75 × 0.2 mm³, inverse resistance ratio (IRR) = R(4K)/R(300K) = 3.2×10³; detailed results in supplementary information (SI) and Ref. [38]) and sample II (s$^{II}$, 1.2 × 0.85 × 0.23 mm³, IRR = 4.0×10⁴, comparable to other high-quality samples [18]). XRD and WDS confirmed the Sm:B stoichiometry of 1:6 (see SI–I(a)). The TCMI technique measures the in-phase ($V'$) and out-of-phase ($V''$) pickup voltages, where $V' \propto$ sample's magnetic susceptibility ($\chi$) and $\sigma(\nu)$ (see SI–II for details).

Figure 1(a) presents the normalized TCMI response $\frac{V'(T)}{\chi(T)} \propto \sigma(\nu, T)$ [37] of s$^{II}$ under 200 mA AC excitation at $\nu = 4$ kHz. The signal exhibits a hump at $T^*$ ~62 K. Below $T^*$, it decreases to a minimum at $T_g$ ~23 K, followed by an increase at $T < T_g$, similar to s$^I$ behavior at $T < T_g = 40$ K [38]. These observations identify three $T$ regimes also for s$^{II}$: (i) $T \geq T^*$ (~62 K) (ii) $T_g$ (~23 K) $\leq T < T^*$, and (iii) $T < T_g$, marked as green, yellow, and blue regions respectively in Fig. 1(a). The inset of Fig. 1(a) shows $V'(v) \propto v^p$ behavior below $T_g$, where $p = 0.464 \pm 0.001$ for $s^I$ [38] and $p = 0.597 \pm 0.002$ for $s^{II}$. In SI–III, this feature is shown at different $T < T_g$ for both samples. Although $T_g$ changes with IRR, both $s^I$ and $s^{II}$ exhibit the characteristic $V'(v) \propto v^{(\sim 0.5)}$ behavior below $T_g$. At $T < T_g = 23$ K, inset of Fig. 1(b) shows the sharp rise in bulk $\rho(T)$ for $s^{II}$. In Fig. 1(b), for $T > T^*$, $\rho(T)$ of $s^{II}$ fits the Hamann function

$$\rho(T) = \rho_{K0}\left[1 - \frac{\ln(T/T_{eff})}{\sqrt{\ln^2(T/T_{eff}) + s(s+1)\pi^2}}\right] + \rho_0 \,[40],$$

with fitting parameters $T_{eff} = 7.31$ K $\pm 0.11$ K, $s = 3.04 \pm 0.01$, $\rho_{K0} = 1.49 \times 10^{-7} \pm 0.05 \times 10^{-7}$ $\Omega$.cm and $\rho_0 = 8.55 \times 10^{-8} \pm 0.36 \times 10^{-8}$ $\Omega$.cm. From the fit, the effective Kondo temperature $T_{eff}$, is quite low (marked as $T_k^S$ in Fig. 1(b) inset). Above $T^*$ (located in $\rho(T)$ via residual analysis, see SI-I(b)), the weak correlation effects result in good agreement with Hamann function. Below $T^*$, the onset of strong correlations results in subtle deviations in $\rho(T)$ from Hamann fit. However, feature at $T^*$ is more clearly identifiable in TCMI measurement (Fig. 1(a)), and in $\chi(T)$ (SI-V)). Below $T_{eff}$, the $\rho(T)$ begins saturating, consistent with earlier reports [2-4,41,42].

The specific heat $C(T)$ behavior of SmB$_6$ follows [43]: $C(T) = \gamma T + \beta_3 T^3 + A T^3 \ln\left(\frac{T}{T_{MFL}}\right) + B T^{-2}$, where $\gamma T$, $\beta_3 T^3$, $A T^3 \ln(T/T_{MFL})$ represent the electronic, phononic, exchange-enhanced paramagnetic contribution, respectively, and $BT^{-2}$ denotes a Schottky anomaly [9,43,44]. The inset of Fig. 1(c) shows $C(T)$ of $s^{II}$ under magnetic fields ($H$) 0 T to 9 T. Figure 1(c) shows the $C(T)$ data fitted with a single curve following the equation $C(T) = \gamma T + \beta_3 T^3 + A T^3 \ln\left(\frac{T}{T_{MFL}}\right) + B T^{-2}$. The fitting parameters are shown in SI IV-(b). However, the single fit done over entire $T$ range makes it difficult to identify the different $T$ regimes where different contributions dominate. To identify distinct $T$ regimes of various contributions, we have done a residual analysis of the $C(T)$ measurement on $s^{II}$. The residual analysis presented in the Fig. 1(d) illustrates the difference between the measured specific heat data, $C_{data}$, and the specific heat values calculated ($C_{fit}$) using the fit parameters obtained from the fitting of $C(T)$ data in Figure 1(c). This is presented as $|C_{data} - C_{fit}|$ vs $T$. The regimes where the $|C_{data} - C_{fit}| \sim 0$ represent regions where the fitting is good. For $T > T^*$, a phonon-dominated behavior is observed. Below $T^*$, as strong correlations emerge (also seen in $\rho(T < T^*)$, Fig. 1(b)), $C(T)$ develops a temperature dependence. Below $T_g$ the $C(T)$ data fits to $C_{Magnetic\ fluctuations} = A\,T^3 ln\left(\frac{T}{T_{MFL}}\right)$ (cyan circles) down to $T_{MFL} \sim 10$ K, indicating diminishing magnetic fluctuations down to $T_{MFL}$. Interestingly, for s$^{II}$ (see SI-V) a history independent bump emerges in $\chi(T)$ between $T^* = 62$ K and $T_{MFL} = 10$ K similar to s$^I$ [38] (and earlier reports [45]), linking it to magnetic fluctuations in SmB$_6$. Below $T_{MFL}$, $C(T)$ fits to $C_{Schottky} = B\,T^{-2}$ (blue circles), confirming Schottky contribution from in-gap states [16,46] within the $\sim 10$ meV bulk gap in SmB$_6$. Only below $T_k^s$, $C(T)$ follows $C_{electronic} = \gamma\,T$, (red circles Fig.1(d), also see SI-IV) with a high $\gamma = 39.2 \pm 1.0$ mJ.K$^{-2}$.mol$^{-1}$ consistent with earlier reports [44]. The green circles indicate the residual for the $C(T)$ data fit with the single equation consisting of all the parameters, which shows excellent fit for the entire $T$ regime. We study the in-gap features in more detail below.

For *I-V* measurements in high-resistance systems like KI, we prefer current control to minimize current fluctuations, which are commonly induced by voltage fluctuations in voltage-controlled *I-V* measurements. Figures 2(a) and 2(b) show the nonlinear *I-V* behavior of s$^{II}$ for $T < T_g$ (~23 K), transitioning to a linear response for $T > T_g$ (see Fig. 2(b) inset). Joule heating effects are negligible (see SI-VI(a)) as the temperature rise is $\ll 0.1$ K with $I \sim 5$ mA during measurements, even at low $T$. The data fits to: $I = I_s\,[exp(eV/k_B T) - 1]$, where $I_s$ is the Schottky saturation current [47]. Figure 2(c) shows *I-V* at 5 K, with the red dashed line representing the fit. The data deviates beyond $I_s(5\,K) = 1.8$ mA, when the excitation energy exceeds a characteristic gap energy ($\varphi_s$) in SmB$_6$. The in-gap, $\varphi_s$, is determined by extracting $I_s(T)$ from *I-V* at different $T$ and fitting to: $I_s(T) = K\,T^2\,exp(\varphi_s/k_B T)$, where $K$ is a constant [47]. Figure 2(d) plots $ln(I_s/T^2)$ vs. $1/T$, yielding $\varphi_s = 2.22 \pm 0.10$ meV for s$^{II}$ ($\varphi_s = 4.20 \pm 0.10$ meV for s$^I$, see SI-VI (b)). An independent estimate of $\phi_s \approx 0.28 \pm 0.12$ meV (s$^{II}$), obtained from the $C(T)$-Schottky anomaly fit parameter $B \propto \phi_s^2$ (see SI-IV(c)), agrees to a value within an order of magnitude of the gap extracted from *I-V*, considering the experimental and modelling limitations (SI-IV(c)). The in-gap $\varphi_s$ is less than the bulk Kondo gap in SmB$_6$ (~10 meV) [4,15,16,21-26]. Note that $\varphi_s$ value is consistent with the location of the Dirac point energy of the linearly dispersing sub-gap surface states $\sim -5$ meV below $E_f$ in SmB$_6$, characteristic of topological Kondo insulator phase [16]. Notably, the *I-V* nonlinearity coincides with $V'(\nu) \propto \nu^{(\sim 0.5)}$ for $T < T_g$ regime, reinforcing its link to surface conduction features. [38].

Figure 3(a) presents for $s^{II}$, the absolute value of the measured $V''$ ($|V''|$) versus $\nu$ at different $T$. The changes in $|V''(\nu)|$ are significant across a ($T$- dependent) characteristic frequency $\nu_0$, where $|V''(\nu_0)| \to 0$. At 45 K and 30 K, for $\nu < \nu_0$, $|V''(\nu)|$ remains featureless (Fig. 3(a)), with $V'(\nu) \propto \nu^2$ (see Fig. 3(b), dark yellow dashed line fit with $\chi^2_{fit\_45K}$= 7.026 × 10$^{-4}$ and $\chi^2_{fit\_30K}$= 8.512 × 10$^{-4}$). For $\nu > \nu_0$, $|V''(\nu)|$ rises sharply, while Fig. 3(b) shows a crossover to $V'(\nu) \propto \nu^p$ behavior with $p \sim 0.4$ - 0.6 (see SI-VII(a) for residual analysis to determine $\nu_0$ in $V'(\nu)$ and similar data for $s^I$, see SI-VII (d) and SI-VIII). Previous studies [37,38] show that a shift in $p$ value from 2 to $\leq 1$ signifies a transformation from the AC field probing the bulk to probing the surface conductivity in a topological material. At low $T$ = 17 K ($T < T_g$), where surface conductivity dominates, viz., $V'(\nu) \propto \nu^{0.6}$ over the entire $\nu$ window (Fig. 3(b)), the highly conducting surface layer effectively shields most of the bulk from the AC electromagnetic field, thereby leaving only a broad minima in $|V''|$ at $\nu_0 \sim 4$ kHz (identified from $d|V''|/d\nu \sim 0$) (see Fig. 3(a)).

The phase difference $\phi(\nu)$ (= $\tan^{-1}\left(\frac{V''(\nu)}{V'(\nu)}\right)$) varies with frequency (see SI-IX), showing changes in the sample's reactive response. To analyse how reactive components influence $|V''(\nu)|$ (Fig. 3(a)), we plot $V''$ vs. $V'$ and compare it with Nyquist plots (see SI-X for plots at different $T$) of AC impedance ($Z''$ vs. $Z'$) simulated using different circuit models [48,49] (see SI-X Figs. 4(a)-(k) and Table-II for the parameter values). Note we match the shape, not the values of $V''(V')$ and $Z''(Z')$, making the circuit component values representative rather than exact. At 294 K (Fig. 4(a)), both curves align with a simple resistance-capacitance (R-C) circuit, with no characteristic $\nu_0$ observed. As $T$ decreases into regime (ii) ($T_g < T < T^*$, Fig. 1(a)), $V''(V')$ curvature completely changes from that at 294 K. At 45 K, the $V''(V')$ curves for $\nu < \nu_0$ (Fig. 4(b)) is not analytically continuous with those at $\nu > \nu_0$ (Fig. 4(c)). At 45 K, in the bulk conductivity-dominated regime ($\nu < \nu_0$, see Fig. 3), a conventional R-C circuit cannot model the Nyquist curve. Instead, an $R_1$-$R_2$ along with a constant phase element (CPE) model the $V''(V')$ curve well (Fig. 4(b)). The CPE impedance, $Z_{CPE} = \frac{1}{Q_0(i\nu)^n}$ [50,51], accounts for non-ideal relaxation, where $Q_0$ is capacitance and $n$ quantifies deviation from ideal C - impedance behavior [50]. To match the $V''(V')$ - $Z''(Z')$ curve in the surface-conduction-dominated regime ($\nu > \nu_0$, see Fig. 4(c)), an additional inductance (L) component is needed, which represents the reactive response of the light high-mobility surface quasiparticles.

Notably, where surface conduction dominates at 2 K ($T < T_g$) in s$^{II}$ (Fig. 4(a) inset), the $V''(V')$ curve is best described with $R - L - CPE$ circuit for all $\nu$. Like the CPE term, previous studies on SmB$_6$ report observing low-$T$ capacitance (~μF), attributed to insulating inclusions with high conducting regions [2]. We observe that CPE is needed for $T \leq 90$ K across all frequencies, however there is an emergent L parameter in the $\nu$ and $T$ regime where surface conduction dominates (see SI-X). Figure 4(d) and (e) shows the temperature evolution of $R_1, R_2$ and $Q_0$, L respectively.

**Discussion:**

Strong correlation, magnetic fluctuations and weak disorder effects modify Drude conductivity [52,53] to, $\sigma(\nu) = \sigma_{Drude}(\nu)[1 - (Q/(k_F \Lambda)^2][1 - (\Lambda/\mathcal{L}_\nu)]$, where $\Lambda = v_F \tau \sim 1$ $nm$ [34,54] is the estimated mean free path for SmB$_6$ with scattering time $\tau \sim 10^{-13}$ s ($T < 10$ K in SmB$_6$) [54], $k_F$ is the Fermi wave vector, a constant $Q \sim O(1)$ and $\mathcal{L}_\nu$ is scattering-free characteristic diffusion length scale of quasiparticles within one period of incident radiation. The $\mathcal{L}_\nu = (D/2\pi\nu)^{1/2}$ where $D = (\Lambda)^2/3\tau \sim 3.3 \times 10^{-6}$ m$^2$s$^{-1}$ is the estimated Diffusion constant of SmB$_6$ in the surface conducting regime, introduces a $\nu^{1/2}$ correction to $\sigma(\nu)$. This correction is consistent with $V'(\nu)$ $(\propto \sigma(\nu)) \sim \nu^{\sim 0.5}$, at all $\nu$ for $T < T_g$, and at $\nu > \nu_0$ for $T > T_g$. Considering protection from scattering in the TI state, we use no $T$ dependence of $\tau$, $\Lambda$ and $D$. $\mathcal{L}_{\nu_0}$ quantifies the phase volume of the highly conducting, low-scattering (topological) component of the fluid present on SmB$_6$ surface. Using $\nu_0(T)$ values, Fig. 4(f) shows an almost diverging nature of $\mathcal{L}_{\nu_0}(T)$ for $T < T_g$ which fits to $A/|T - T_c|^\beta$, where $A = 35.96 \pm 6.02$ μm.K, $\beta \sim 0.50 \pm 0.05$ and $T_c \sim 5 \pm 2$ K. This $T_c \sim T_k^s$, seen in $C(T)$ (Fig. 1(d)) and in $\rho(T)$ (Fig.1(b)). Thus, at low $T \leq T_k^s$, the region with high-mobility surface-dominated phase grows. What is the significance of $T_k^s$? Also $V'(\nu)$ $(\propto \sigma(\nu)) \sim \nu^{\sim 0.5}$, suggests role of magnetic fluctuations — where does it arise from?

Our results align with the Kondo breakdown (KB) model for TKIs [55,56]. In SmB$_6$, surface Sm vacancies [34,55] release carriers that weaken Kondo screening, lowering the surface Kondo temperature to $T_k^s \sim O(T_K/10)$, where $T_k$ is the bulk Kondo temperature (Kondo gap ~10 meV [4,21] $\equiv$ 116 K ). The KB-driven effect releases surface magnetic fluctuations from $T < T^*$, which is marked by a cusp in $V'(T)$(Fig. 1(a)) , deviations in $\rho(T)$ below $T^*$ (Fig.

1(b)) due to fluctuation induced scattering and non-phononic contribution in $C(T < T^*)$ (Fig. 1(c)). In Figs. 4(d) and 4(e), $R_1$ captures bulk resistance, while $R_2$ is associated with charged heavy carriers released on the surface due to broken Kondo state. Strong correlations and localization of these carriers cause $R_2$ to sharply increase with reducing $T$ below $T_g$. The CPE ($\equiv Q_0$) captures the distributed capacitance from strong correlations and spatially inhomogeneous relaxation effects, showing a peak below $T^*$. $Q_0$ saturates below $T_k^s$ - marking the emergence of a topological surface state. The L, negligible at high $T$, rises rapidly below $T_g$, indicating the growing kinetic inductance of high-mobility surface carriers in the TI state ($T < T_k^s$). While below $T_k^s$ [2-4,41,42] the high conducting surface state covers most of the sample (as $\mathcal{L}_{\nu_0}(T)$ becomes large), this sheath of lighter Dirac fluid starts developing from $T < T_g$ itself. Note below $T_k^s$, $R_2$ contribution while still present, its rate of increase slows down. Thus, light ($\equiv L$) and heavy ($\equiv R_2$) fluids seem to coexist to determine the surface conducting properties below $T_k^s$.

As $T$ decreases, KB induced carriers progressively resume Kondo screening of the surface magnetic fluctuations [55,56]. In $C(T)$ (Figs.1(c),1(d)), these fluctuations subside between $T_g$ and $T_{MFL}$. As surface Kondo screening revives, $\rho(T)$ rises rapidly below $T_g$ (Fig. 1(b)). Concomitantly, suppression of magnetic fluctuations at low $T$ also revives time-reversal symmetry, thereby stabilizing the topological surface state [55]. In fact below $T_g$ although the bulk is Kondo-screened, the high mobility light (topological) quasiparticle fluid phase expands (diverging $\mathcal{L}_{\nu_0}(T)$) with predominant $V'(\nu)$ ($\propto \sigma(\nu)$) $\sim \nu^{\sim 0.5}$ behavior, characteristic of high-conducting region on the surface. Note, the resemblance of $\mathcal{L}_{\nu_0}(T)$ to critical phenomena like behavior warrants future investigations. The emergent surface state recovers below $T_s^k \approx 7$ K, seen from the saturating trend in $\rho(T)$ below $T_s^k$ (Fig.1(b)), and also from $\mathcal{L}_{\nu_0}(T)$ fit. The lighter Dirac-like quasiparticles below $T_g$ are excited across the gap $\phi_s$ (recall Fig. 2).

Not all quasiparticles released on surface by broken Kondo effect take part in the Kondo screening of Sm moments in the bulk. Some of these strongly correlated particles remain free on surface down to low $T$, where they are detected as particles with high $\gamma$ in $C(T)$. Therefore, the surface of SmB$_6$ has a two fluid like phase with, (i) a heavy, defect generated strongly correlated quasiparticle, and (ii) a light (Dirac-like), high mobility quasiparticle. Although the lighter phase grows with reducing $T$ ($< T_s^k$), (divergence of $\mathcal{L}_{\nu_0}(T)$), the two fluid phase on the surface of SmB$_6$ survives down to low $T$. High effective mass of heavy quasiparticles in the fluid, for $T \leq 90$ K produces a non-ideal capacitive (CPE) response (Figs. 4(b), (c) and SI-X). Effective AC field screening by the lighter quasiparticles (below $T_g$), allows their reactive behavior to be modelled well with an inductive (L) component. In fact for $\nu < \nu_0$ to $> \nu_0$ at $T > T_g$ or at all $\nu$, below $T_g$, the two-fluid state in SmB$_6$ shows a crossover from a purely CPE to CPE plus L response for capturing the dynamics of mobile and sluggish surface quasiparticles (Figs. 4(a)-(c)).

**Conclusion:**

In conclusion, rather than hosting purely topological Dirac-like particles, our study reveals that SmB$_6$ surface hosts a disorder-driven two-fluid like phase with both heavy and light quasiparticles. Understanding SmB$_6$ is crucial for exploring how strong correlations and disorder affect band topology and lead to new emergent quantum phases which is of fundamental importance in condensed matter physics.

**Acknowledgements:** The authors thank Priscila F. S. Rosa (LANL) and Dr. Amit Jash (Weizmann Institute) for support. S. S. B acknowledges infrastructure and funding from the DST-SERB SUPRA, DST-AMT, Govt. of India (GOI) programs, and IIT Kanpur. S.G thanks CSIR, India for financial support. S.P acknowledges the PMRF Scheme support from Ministry of HRD, GOI.

**Note:** Sayantan Ghosh and Sugata Paul have equal contributions.

**REFERENCES**

[1] J. G. Checkelsky, B. A. Bernevig, P. Coleman, Q. Si, and S. Paschen, Nature Reviews Materials **9**, 509 (2024).
[2] M. Dzero, J. Xia, V. Galitski, and P. Coleman, Annual Review of Condensed Matter Physics **7**, 249 (2016).
[3] Z. F. Priscila F. S. Rosa, in *Rare-Earth Borides*, edited by D. S. Inosov (Jenny Stanford Publishing Pte. Ltd., Singapore, 2021), pp. 817.
[4] L. Li, K. Sun, C. Kurdak, and J. W. Allen, Nature Reviews Physics **2**, 463 (2020).
[5] J. W. Allen, B. Batlogg, and P. Wachter, Physical Review B **20**, 4807 (1979).
[6] R. M. Martin and J. W. Allen, Journal of Applied Physics **50**, 7561 (1979).
[7] Z. Fisk, J. L. Sarrao, S. L. Cooper, P. Nyhus, G. S. Boebinger, A. Passner, and P. C. Canfield, Physica B: Condensed Matter **223-224**, 409 (1996).
[8] H. Tsunetsugu, M. Sigrist, and K. Ueda, Reviews of Modern Physics **69**, 809 (1997).
[9] P. S. Riseborough, Advances in Physics **49**, 257 (2000).
[10] M. Dzero, K. Sun, V. Galitski, and P. Coleman, Physical Review Letters **104**, 106408 (2010).
[11] M. Dzero, K. Sun, P. Coleman, and V. Galitski, Physical Review B **85**, 045130 (2012).
[12] S. Wolgast, Ç. Kurdak, K. Sun, J. W. Allen, D.-J. Kim, and Z. Fisk, Physical Review B **88**, 180405 (2013).
[13] P. Syers, D. Kim, M. S. Fuhrer, and J. Paglione, Physical Review Letters **114**, 096601 (2015).
[14] Y. S. Eo, A. Rakoski, J. Lucien, D. Mihaliov, Ç. Kurdak, P. F. S. Rosa, and Z. Fisk, Proc. Natl. Acad. Sci. U.S.A **116**, 12638 (2019).
[15] M. Neupane *et al.*, Nature Communications **4**, 2991 (2013).
[16] H. Pirie *et al.*, Nature Physics **16**, 52 (2020).
[17] G. Li *et al.*, Science **346**, 1208 (2014).
[18] B. S. Tan *et al.*, Science **349**, 287 (2015).
[19] M. Hartstein, H. Liu, Y.-T. Hsu, B. S. Tan, M. Ciomaga Hatnean, G. Balakrishnan, and S. E. Sebastian, iScience **23**, 101632 (2020).
[20] M. Hartstein *et al.*, Nature Physics **14**, 166 (2018).
[21] H. Miyazaki, T. Hajiri, T. Ito, S. Kunii, and S.-i. Kimura, Physical Review B **86**, 075105 (2012).
[22] J. Jiang *et al.*, Nature Communications **4**, 3010 (2013).
[23] E. Frantzeskakis *et al.*, Physical Review X **3**, 041024 (2013).
[24] N. Xu *et al.*, Physical Review B **90**, 085148 (2014).
[25] W. T. Fuhrman *et al.*, Physical Review Letters **114**, 036401 (2015).
[26] X. Zhang, N. P. Butch, P. Syers, S. Ziemak, R. L. Greene, and J. Paglione, Physical Review X **3**, 011011 (2013).
[27] P. S. Riseborough and Z. Fisk, Physical Review B **96**, 195122 (2017).
[28] G. Baskaran, arXiv:1507.03477., (2015).
[29] J. Knolle and N. R. Cooper, Physical Review Letters **118**, 096604 (2017).
[30] O. Erten, P.-Y. Chang, P. Coleman, and A. M. Tsvelik, Physical Review Letters **119**, 057603 (2017).
[31] D. Chowdhury, I. Sodemann, and T. Senthil, Nature Communications **9**, 1766 (2018).
[32] S. Rößler, L. Jiao, D. J. Kim, S. Seiro, K. Rasim, F. Steglich, L. H. Tjeng, Z. Fisk, and S. Wirth, Philosophical Magazine **96**, 3262 (2016).
[33] S. Sen, N. S. Vidhyadhiraja, E. Miranda, V. Dobrosavljević, and W. Ku, Physical Review Research **2**, 033370 (2020).
[34] C. E. Matt *et al.*, Physical Review B **101**, 085142 (2020).
[35] M. E. Boulanger *et al.*, Physical Review B **97**, 245141 (2018).
[36] A. T. Fiory, A. F. Hebard, P. M. Mankiewich, and R. E. Howard, Appl. Phys. Lett. **52,** 2165–2167 (1988).

[37] A. Jash, K. Nath, T. R. Devidas, A. Bharathi, and S. S. Banerjee, Physical Review Applied **12**, 014056 (2019).
[38] S. Ghosh, S. Paul, A. Jash, Z. Fisk, and S. S. Banerjee, Physical Review B **108**, 205101 (2023).
[39] A. Jash, S. Ghosh, A. Bharathi, and S. S. Banerjee, Bulletin of Materials Science **45**, 17 (2022).
[40] D. R. Hamann, Physical Review **158**, 570 (1967).
[41] A. Menth, E. Buehler, and T. H. Geballe, Physical Review Letters **22**, 295 (1969).
[42] J. C. Cooley, M. C. Aronson, Z. Fisk, and P. C. Canfield, Physical Review Letters **74**, 1629 (1995).
[43] W. A. Phelan, S. M. Koohpayeh, P. Cottingham, J. W. Freeland, J. C. Leiner, C. L. Broholm, and T. M. McQueen, Physical Review X **4**, 031012 (2014).
[44] J. Stankiewicz, M. Evangelisti, P. F. S. Rosa, P. Schlottmann, and Z. Fisk, Physical Review B **99**, 045138 (2019).
[45] P. K. Biswas *et al.*, Physical Review B **89**, 161107 (2014).
[46] K. Flachbart, K. Gloos, E. Konovalova, Y. Paderno, M. Reiffers, P. Samuely, and P. Švec, Physical Review B **64**, 085104 (2001).
[47] D. V. Averyanov *et al.*, Sci Rep **6**, 25980 (2016).
[48] Y. Araki and J. i. Ieda, Journal of the Physical Society of Japan **92**, 074705 (2023).
[49] J. Dong, V. Juričić, and B. Roy, Physical Review Research **3**, 023056 (2021).
[50] J. R. Macdonald, Annals of Biomedical Engineering **20**, 289 (1992).
[51] J. Stankiewicz, P. Schlottmann, J. Blasco, M. C. Hatnean, and G. Balakrishnan, Materials Research Bulletin **159**, 112105 (2023).
[52] N. F. Mott and M. Kaveh, Advances in Physics **34**, 329 (1985).
[53] K. Lee, A. J. Heeger, and Y. Cao, Physical Review B **48**, 14884 (1993).
[54] J. Zhang, J. Yong, I. Takeuchi, R. L. Greene, and R. D. Averitt, Physical Review B **97**, 155119 (2018).
[55] V. Alexandrov, P. Coleman, and O. Erten, Physical Review Letters **114**, 177202 (2015).
[56] O. Erten, P. Ghaemi, and P. Coleman, Physical Review Letters **116**, 046403 (2016).

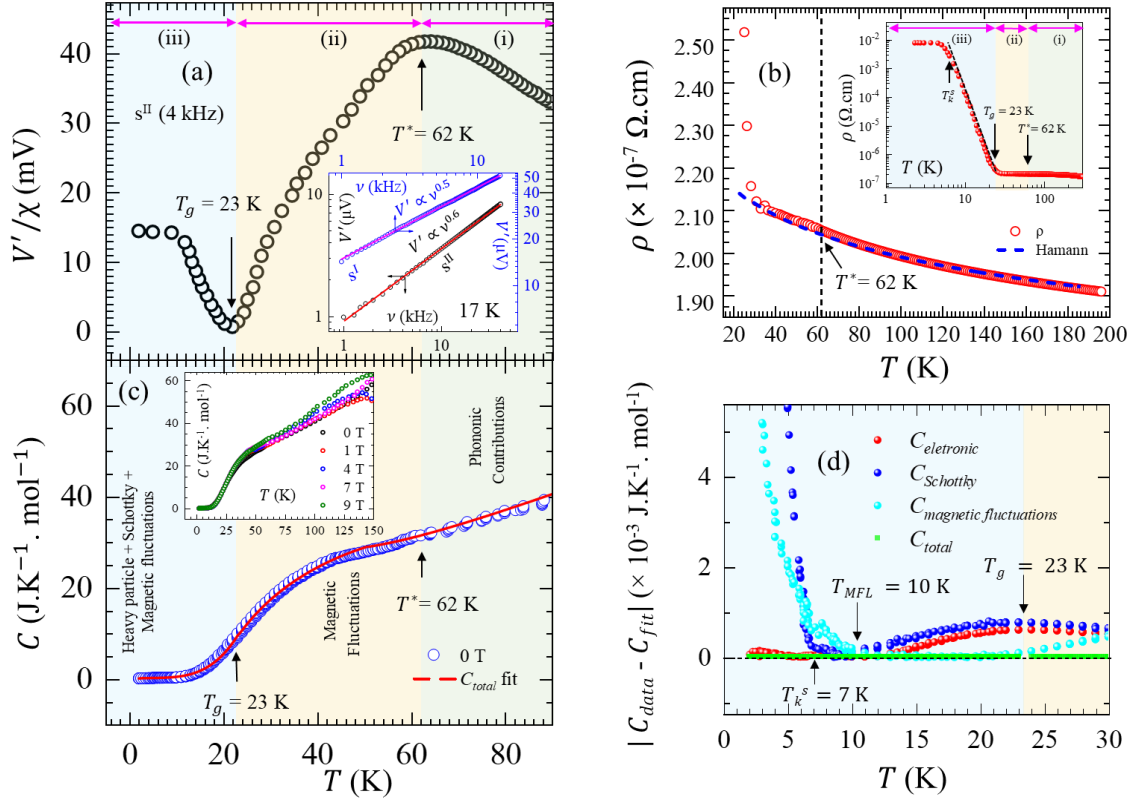

**Fig.1.** (a) $\frac{V'(T)}{\chi(T)}$ of s$^{II}$ for 200 mA AC excitation at $\nu = 4$ kHz, showing three temperature regimes (i) (green), (ii) (yellow) and (iii) (blue), separated by $T^* \sim 62$ K and $T_g \sim 23$ K (magenta arrows). Inset: $V'(\nu)$ at 17 K for s$^{II}$ (black circles) with red fit $V'(\nu) \propto \nu^{0.6}$ and s$^{I}$ (blue circles) with magenta fit of $V'(\nu) \propto \nu^{0.5}$. (b) Inset: $\rho(T)$ for s$^{II}$ ($I = 1$ mA), dashed line is guide to eye showing deviation below $T_k^s$. Main panel: Hamann function fit (blue dashed line) to $\rho(T)$ up to 200 K. (c) Inset: specific heat ($C(T)$) behavior of s$^{II}$ at different magnetic fields. Main panel: $C$ vs $T$ for 0 T with fit of the whole equation $C(T) = \gamma T + \beta_3 T^3 + A T^3 \ln\left(\frac{T}{T_{MFL}}\right) + B T^{-2}$. (d) Residual analysis of $C(T)$ fits using different fitting function for different $T$ regimes. Red, blue and cyan circles represent residual of fitting functions of the terms $C_{electronic} = \gamma T$, $C_{Schottky} = B T^{-2}$ and $C_{Magnetic\ fluctuations} = A T^3 \ln\left(\frac{T}{T_{MFL}}\right)$ respectively (see main text for details). Green circles show residual analysis of the whole equation used in Fig. 1(c).

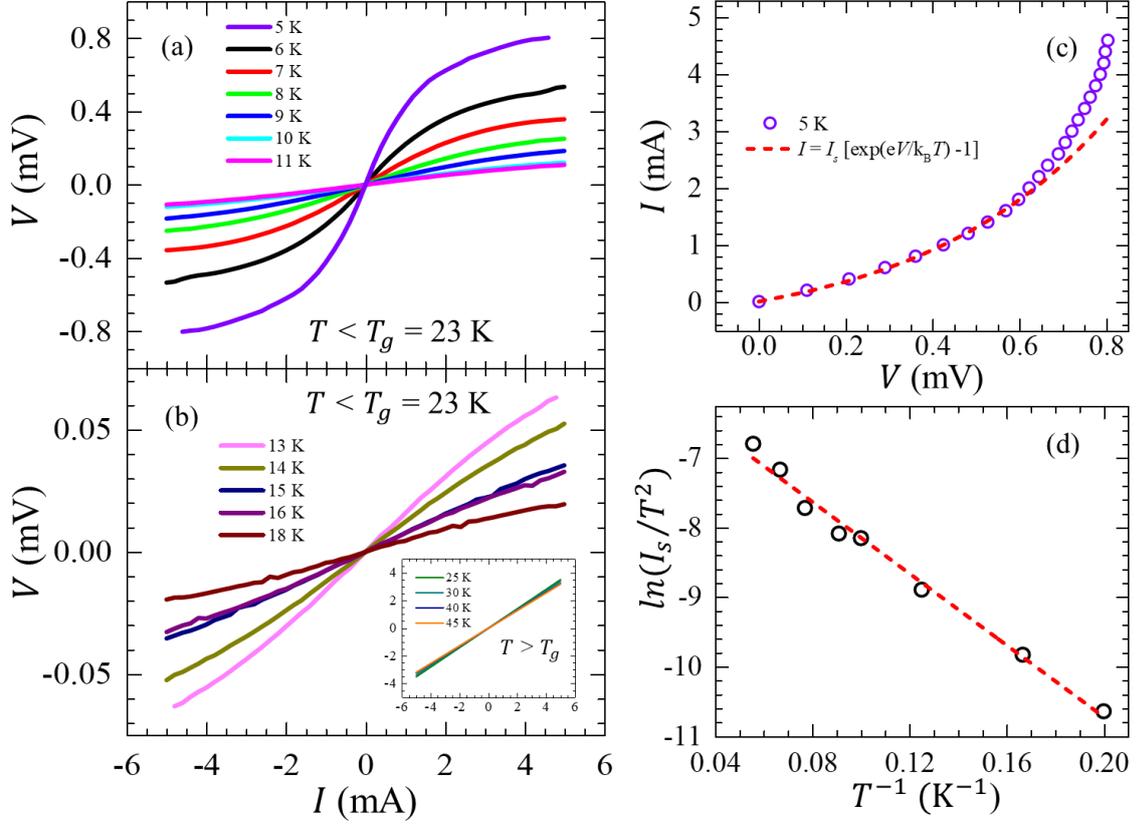

**Fig.2.** (a) and (b) Voltage ($V$) vs applied current ($I$) of SmB$_6$ s$^{II}$ at different $T < T_g$ (See text for details) (b) Inset shows the linear $I - V$ response for $T > T_g$ (regime ii). (c) $I - V$ at 5 K fitted (red dashed line) with $I = I_s[exp\left(\frac{eV}{k_BT}\right) - 1]$. (d) $ln(\frac{I_s}{T^2})$ vs $1/T$ plot with linear fit (red dashed line).

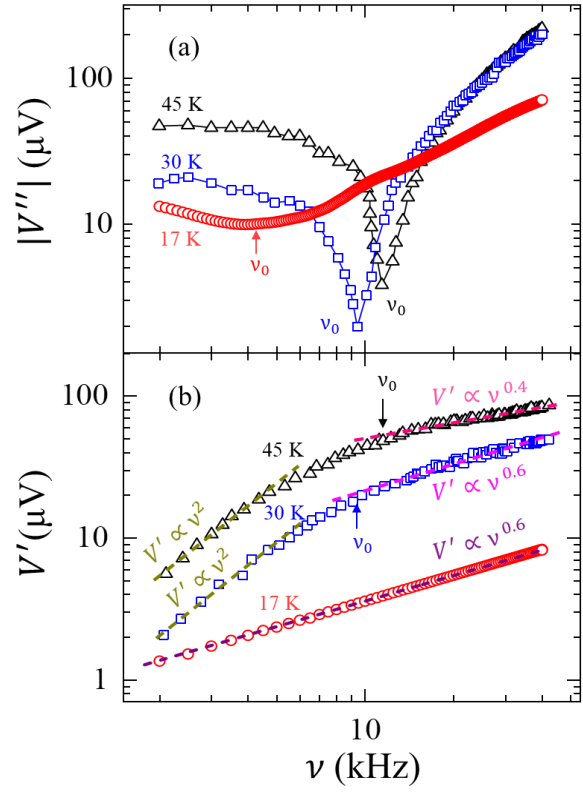

**Fig.3.** (a) $|V''|$ vs $\nu$ and (b) $V'$ vs $\nu$ at 17 K, 30 K and 45 K. Wine, magenta and pink dashed lines represent $V'(\nu) \propto \nu^{0.4-0.6}$ behavior. Dark yellow dashed lines represent $V'(\nu) \propto \nu^2$.

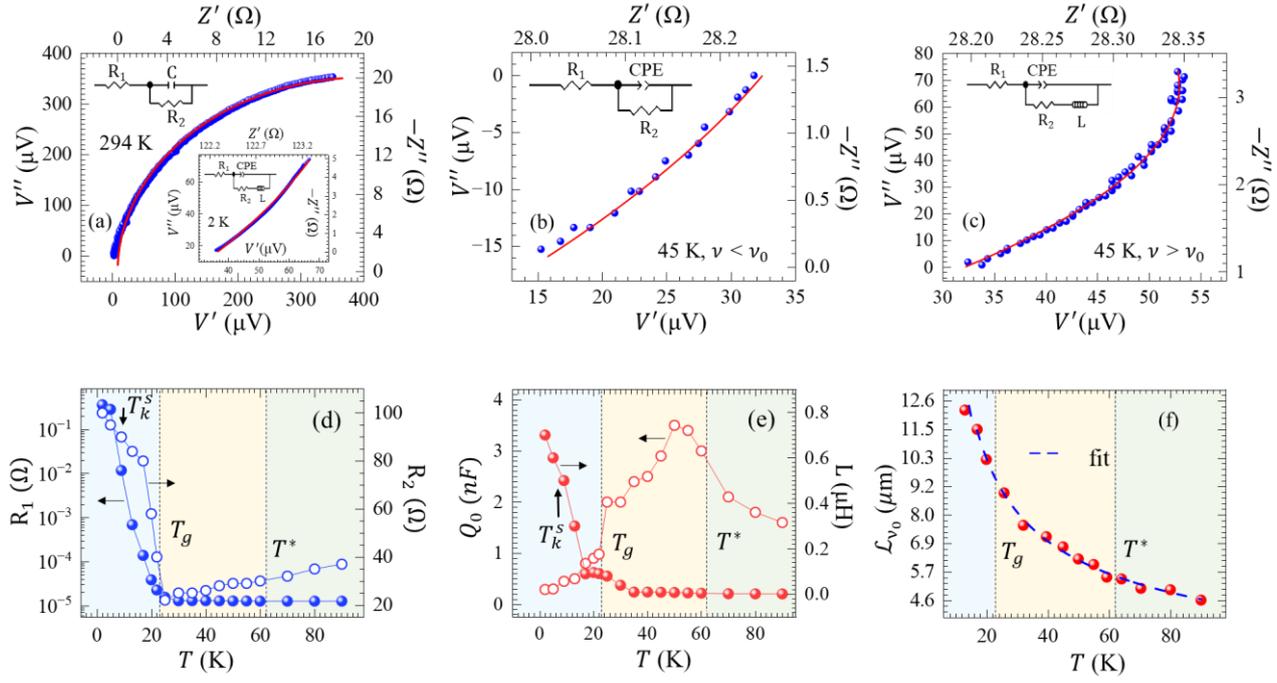

**Fig.4**. Nyquist analysis at (a) 294 K (inset: 2 K), (b) 45 K, $\nu < \nu_0$ (c) 45 K, $\nu > \nu_0$ with corresponding equivalent circuit models. TCMI data $V''(V')$ as blue spheres and simulated equivalent electrical impedance $-Z''(Z')$ in red curve. (d) $R_1$, $R_2$ vs $T$ and (e) $Q_0$, $L$ vs $T$. (f) $\mathcal{L}_{\nu_0}$ vs $T$ behavior.

# Low Temperature Two Fluid State in SmB$_6$

Sayantan Ghosh[1], Sugata Paul[1], Tamoghna Chattoraj[1], Ritesh Kumar[1], Zachary Fisk[2], S. S. Banerjee[1,†]

[1]*Indian Institute of Technology Kanpur, Kanpur, Uttar Pradesh 208016, India.*
[2]*Department of Physics and Astronomy, University of California at Irvine, Irvine, CA 92697, USA.*

Corresponding author email : [†]satyajit@iitk.ac.in

## SECTION I:

### (a) Structural and chemical characterization of SmB$_6$ single crystal s$^{II}$

Fig. 1(a) shows the XRD pattern of our SmB$_6$ single crystal s$^{II}$. We obtained four peaks i.e. (100), (200), (300) and (400) at $2\theta$ positions around 21.49°, 43.9°, 68.02° and 96.61° respectively. These peaks were verified by comparing with the PCD file data for standard powder XRD pattern of SmB$_6$ (refer to Pearson's crystal database (PCD) file no. 1126045).

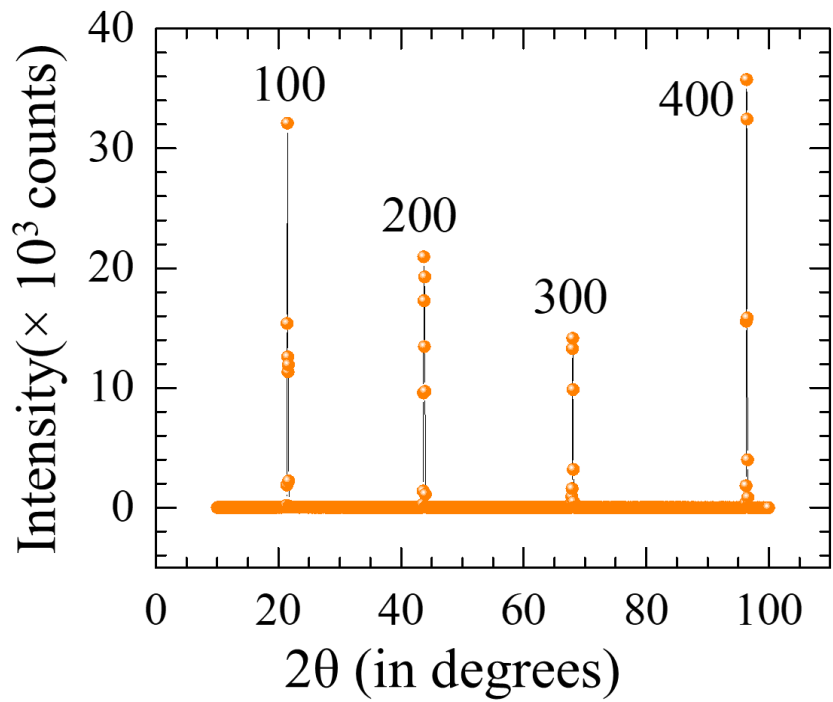

**Fig. 1(a).  XRD of SmB$_6$ single crystal s$^{II}$**

Absence of any additional impurity phases in the XRD pattern ensured that the sample is single phased and a single crystal with growth direction along (100) crystal plane. The Bragg's diffraction condition is given by:

$$2d\sin\theta = n\lambda \text{ ; where } d = a/\sqrt{h^2 + k^2 + l^2}.$$

Here $d$ is the separation between the crystal planes with $a$ being the lattice constant for cubic lattice, $(h\ k\ l)$ are the Miller indices of the plane and $n$ is the order of diffraction. Using $\lambda = 0.15406$ nm as the standard X-ray wavelength (Cu, $K_{\alpha 1}$ line) for the (100) peak, we obtain $a = 0.413$ nm. This value exactly matches with the previously published data for $SmB_6$ [1,2]. Also, there are no other XRD peaks from Aluminium (Al) observed in Fig. 1(a) which confirms the absence of any Aluminium flux present in the bulk of our $SmB_6$ single crystal.

The chemical analysis of the $SmB_6$ single crystal was done using JEOL JXA-8230 Electron Probe Micro Analyser (EPMA) instrument by Wavelength-dispersive X-ray spectroscopy (WDS) measurement. The spectral resolution of WDS is much higher than EDS and hence effective in detecting light elements like Boron (B) present in the sample. The WDS results of $SmB_6$ crystal, as shown in Table-I below, demonstrates the average (calculated from 15 different points on the sample) atomic percentage of Sm and B present in the sample i.e. Sm : B to be close to 1:6. This is consistent with the stoichiometric composition of $SmB_6$. Furthermore, no presence of Aluminium (Al) or any other significant impurity element has been found from the WDS study, proving the $SmB_6$ single crystal to be pure and without any remnant Al percentage in the Al flux grown $SmB_6$ sample.

**Table I: WDS results of $SmB_6$ single crystal**

| Sm (atom %) | B (atom %) | B/Sm |
|---|---|---|
| 14.43 | 85.57 | 5.93 * |

*Average value from 15 different points on the sample.

## (b) Residual analysis to determine $T^*$ location in $\rho(T)$ for $s^{II}$ SmB$_6$

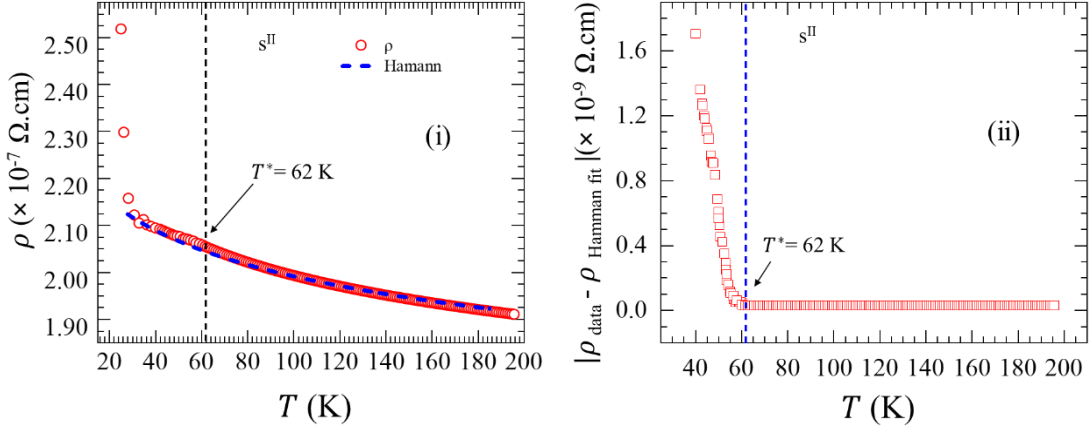

**Fig. 1(b).** (i) $\rho(T)$ data of SmB$_6$ $s^{II}$ up to 200 K with Hamann fit (blue dashed line). Fit deviates below $T^* = 62$ K (black dashed line). (ii) Residual plot of Hamann fit up to 200K for SmB$_6$ $s^{II}$.

Figure 1(b)-(i) above shows the Hamann function fit to the $\rho(T)$ data of SmB$_6$ up to 200 K. We have fitted the $\rho(T)$ data using the Hamann function $\rho(T) = \rho_{K0}\left[1 - \frac{\ln(T/T_{eff})}{\sqrt{\ln^2(T/T_{eff}) + s(s+1)\pi^2}}\right] + \rho_0$, with fitting parameters $T_{eff} = 7.31$ K $\pm$ 0.11 K, $s = 3.04 \pm 0.01$, $\rho_{K0} = 1.49 \times 10^{-7} \pm 0.05 \times 10^{-7}$ $\Omega$.cm and $\rho_0 = 8.55 \times 10^{-8} \pm 0.36 \times 10^{-8}$ $\Omega$.cm. The experimental $\rho(T)$ data gives us an estimate of the order of the parameters $\rho_{K0}$ and $\rho_0$. The $s$ is the effective spin of the Sm moment which is $\sim 3$.

The fit shows from $T^* = 62$ K, the Hamman function diverges from the $\rho(T)$ data as strong correlation starts emerging in the system causing a deviation from the weakly interacting behavior. Figure 1(b)-(ii) above shows the residual analysis $|\rho(T)_{data} - \rho(T)_{Hamann}|$, where $\rho(T)_{Hamann}$ is the $\rho(T)$ values calculated using the fit parameter in the Hamann function. The residue shows a nearly zero value from 200 K down to 62 K and a smooth deviation from zero below $T^* = 62$ K suggesting that the Hamann function is a good fit to the $\rho(T)$ data from high $T$ down to $T^*$.

The signature of $T^*$ in $\rho(T)$ does not seem much prominent, however, we also want to emphasize here that the $T^*$ location is not identified only from $\rho(T)$ behavior, but also from the cusp like feature at $T^*$ in TCMI measurements (Fig 1(a) in main paper), in DC susceptibility

(Fig 5 inset in supplementary section V) and from specific heat (Fig.1(c) of main paper). All these different experiments consistently identify a similar range of $T^* = 62$ K.

# SECTION II:
## TCMI measurement technique

In the two-coil mutual inductance (TCMI) technique (Fig. 2(a) below), the sample is kept between two coils of closely matched parameters. A time-varying AC magnetic field is created by an alternating current ($I_{ac}$) at frequency $\nu$ sent in the excitation coil (Fig. 2(a)). This AC excitation magnetic field is experienced by the sample placed between the two coils. This AC field induces current inside the sample which in turn induces voltage in the pick-up coil (Fig. 2(a)). We measure the pick-up voltage induced by the sample by a Stanford SR830 DSP lock-in amplifier. To reduce the pick-up signal due to stray fields around the sample, we use a 1.5 mm thick oxygen-free high conductivity (OFHC) Copper sheet (with an insulating coat) with a hole of diameter 0.75 mm placed on top of the excitation coil and the sample is placed just above the hole [3-6] (see Fig. 2(a) schematic of the two coil mutual inductance setup). The thickness of the OFHC copper sheet is kept much larger than the skin depth of the lowest AC frequency we use in our measurements. Thus, the thick copper sheet allows for minimal stray field coupling between the two coils (excitation and pickup coil). The only coupling between the two coils is via the $SmB_6$ sample placed over the hole in the copper sheet,

As shown in the schematic Fig. 2(b) below, at low frequencies ($\nu$) of $I_{ac}$ excitation current, the magnetic field of the excitation coil (see left hand schematic in Fig. 2(b) below) induces screening currents ($I_{induced}$) all across the bulk of the sample (see left image in Fig. 2(b), while at high $\nu$ these $I_{induced}$ are induced predominantly on the surface of the sample (see right image in Fig. 2(b) below). These $I_{induced}(\nu)$ generates an oscillating field $h_{ac}$ from the sample which is picked up by the pickup coil. The amplitude of the oscillating magnetic field, $|h_{ac}| \propto I_{induced}(\nu) \propto \sigma(\nu)$, as $I_{induced}(\nu)$ depends directly on the AC conductivity of the material, $\sigma(\nu)$, which controls the amount of $I_{induced}(\nu)$ generated in the material.

The pick-up voltage $V = V' + iV''$, where $V'$ and $V''$ are the in-phase (real) and out-of-phase (imaginary) component of the induced voltage ($V$) respectively. The in-phase component ($V'$) corresponds to the electrical conductivity of the material ($\sigma$) in this technique, as $V' \approx -\mu_0 N k \zeta \chi \left(\frac{dh_{ac}}{dt}\right)$, where $\mu_0$ is the permeability of free space, $N$ is the number of turns of the pick-up coil, $k$ is the geometric filling factor, $\zeta$ is factor depending on the geometry of the coils like the cross sectional area of the coil (both coils have same cross-sectional area) and $\chi$ is the AC susceptibility of the sample [3]. In our paper we usually plot $V'/\chi$ as its proportional to $\sigma(\nu)$. Note that, for all our measurements we have carefully subtracted the background response (without sample) from the total response (sample + background) to obtain the response of only the sample itself. For $SmB_6$ we want to mention that due to the low value of $\chi$ (section-V below), dividing $V'$ by $\chi$ doesn't change the behaviour of $V'$.

COMSOL simulations with current of 200 mA in the excitation coil, show that the thin Cu sheet significantly reduces the stray field from outside the hole and helps concentrate the AC magnetic field directly on the sample which is kept on top of the hole (Fig. 2(c)). The stray fields are shielded even further in our experiment as we use a thick Cu sheet. We see from the Fig. 2(c) that with 200 mA in the AC excitation coil we get a field of ~ 260 Oe focussed on the sample placed above the hole in the OFHC copper sheet.

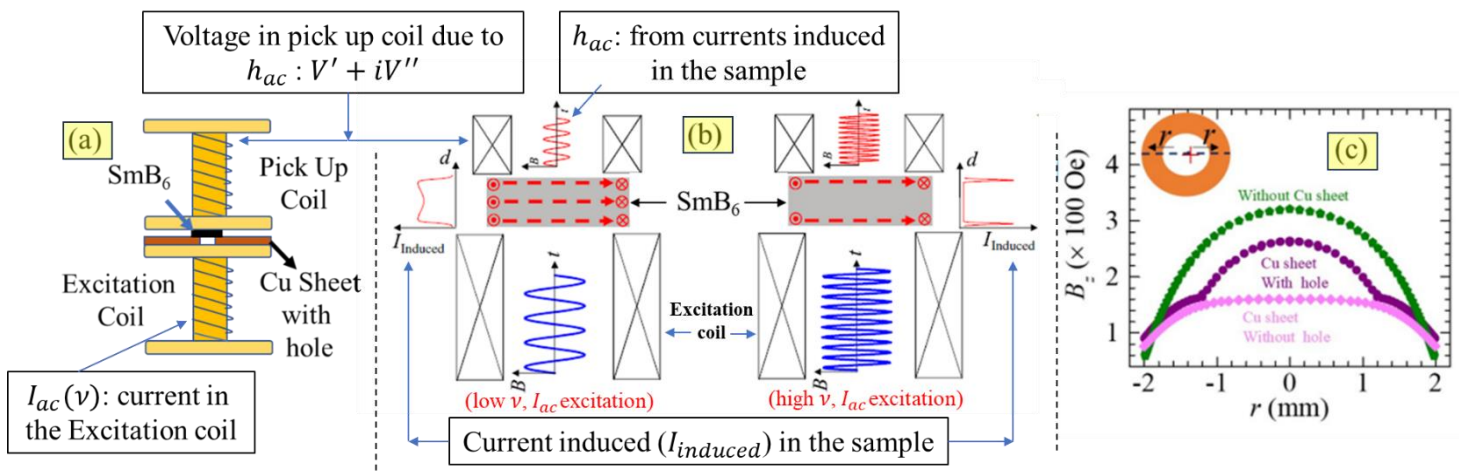

**Fig. 2 (a). Schematic of the TCMI setup [3]. (b) Schematic of the skin-depth penetration for low and high excitation frequency [4]. (c) COMSOL simulation of the magnetic field with and without the Cu sheet on top of the excitation coil [4].**

## Section III:

$V'(\nu) \propto \nu^p, p < 1$ response of SmB$_6$ below $T_g$

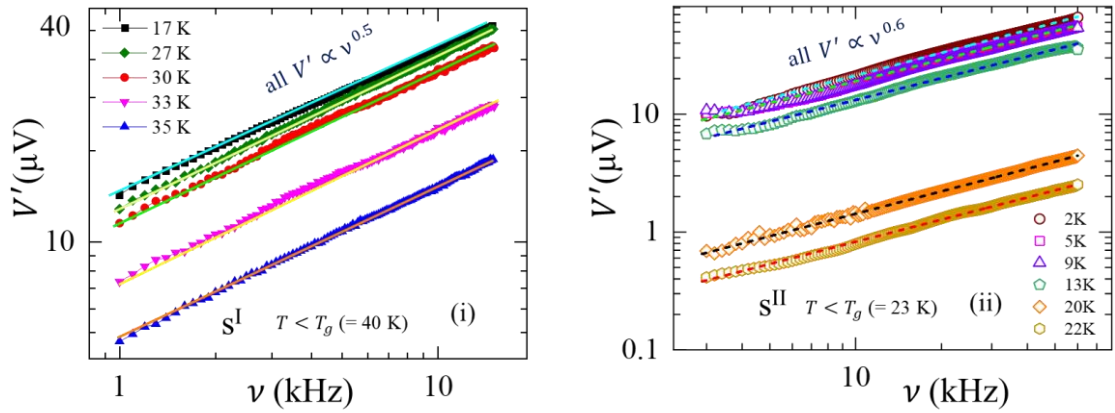

**Figure 3:** $V'(\nu) \propto \nu^p, p < 1$ response of SmB$_6$ below $T_g$. (i) $V'(\nu)$ data showing $V'(\nu) \propto \nu^{0.5}$ behavior for s$^I$ at $T < T_g$ (= 40 K). (ii) $V'(\nu)$ data showing $V'(\nu) \propto \nu^{0.6}$ behaviour for s$^{II}$ at $T < T_g$ (= 23 K).

Figure 3(i) above shows the $V'(\nu) \propto \nu^{0.5}$ behaviour of SmB$_6$ (s$^I$) observed at different temperatures (closed symbols, 17 K, 27 K, 30 K, 33 K and 35 K) below $T_g$ (= 40 K). At each temperature, the $V'(\nu)$ signal is obtained at a constant excitation current amplitude $I_0$ = 200 mA. Similarly, Fig. 3(ii) shows the $V'(\nu) \propto \nu^{0.6}$ response of SmB$_6$ (s$^{II}$) for $T$ = 2 K, 5 K, 9 K, 13 K, 20 K and 22 K which are all below $T_g$ (= 23 K) for s$^{II}$. This strongly supports that the unique $V'(\nu) \propto \nu^p$ (where $p < 1$) behaviour of SmB$_6$ is present in the entire $T < T_g$ regime irrespective of the samples (both s$^I$ and s$^{II}$).

## SECTION IV:

### (a) Specific heat ($C$) response of SmB$_6$ s$^{II}$ at different magnetic fields

Fig. 4(a) shows the $C/T^3$ vs $T$ response of SmB$_6$ s$^{II}$ for 1 T (Fig. 4(a)-(i)), 4 T (Fig. 4(a)-(ii)) 7 T (4(a)-(iii)) and 9 T (Fig. 4(a)-(iv)) magnetic field. $C/T^3 = \gamma/T^2 + \beta_3 + A\ln(T/T_{MFL}) + BT^{-5}$. The first and second terms are standard electronic and lattice contributions to specific heat respectively. The third term is typically associated with an exchange enhanced paramagnetic metal, and $BT^{-2}$ is related to Schottky anomaly present inside SmB$_6$ [7]. Note that the nature of $C/T^3$ vs $T$ does not vary by applying magnetic field. The Schottky behaviour gradually gets suppressed as we increase the applied field, showing the $BT^{-5}$ curve fit (blue solid line) to deviate slightly from the data. The fitting parameters for the fits of the figures with the different magnetic field values from 0 T to 9 T (see the fit for 0 T in main manuscript Fig. 1(c) and Fig. 1(d)) are shown in Table-I below.

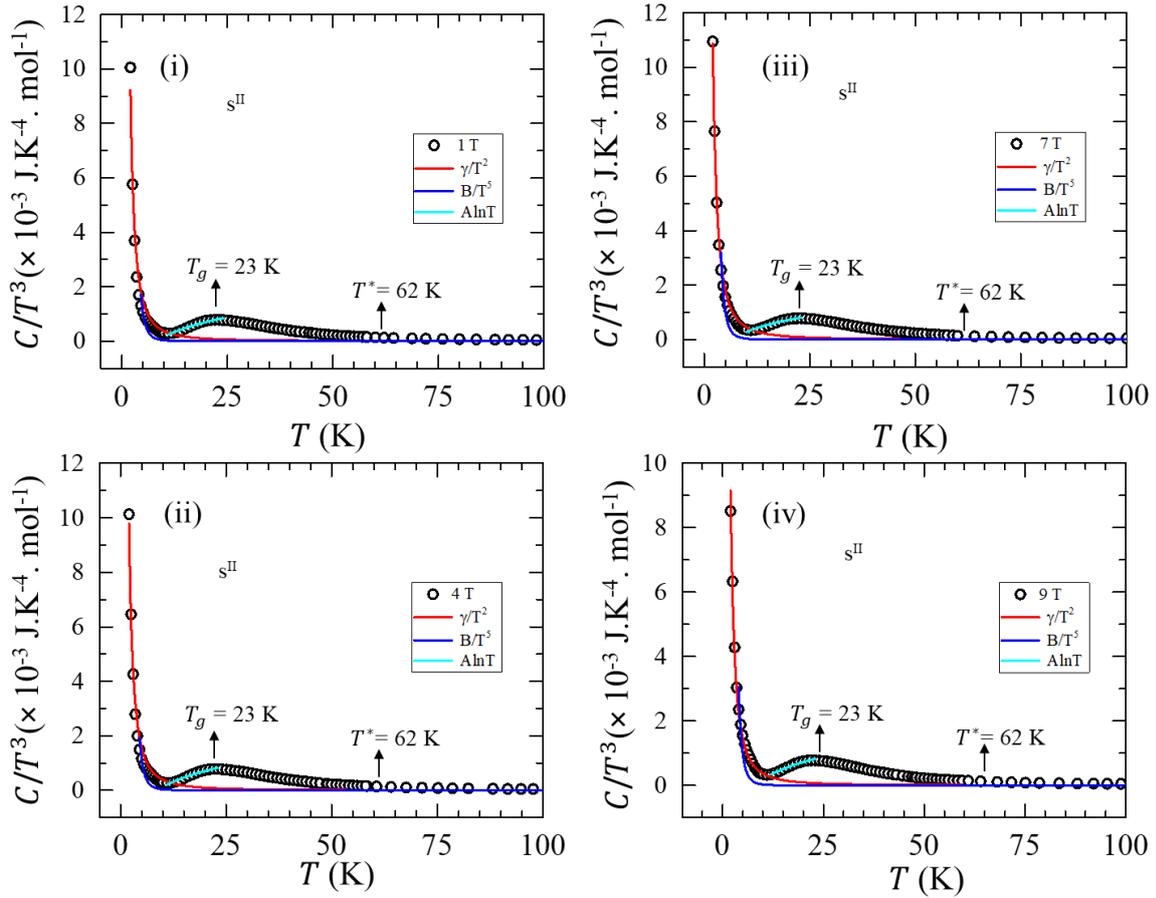

**Fig. 4(a) $C/T^3$ vs $T$ plot for 1 T, 4 T, 7 T and 9 T magnetic field on s$^{II}$**

## Table-I

| Applied field (T) | $\gamma$ (fit) mJ.K$^{-2}$mol$^{-1}$ | B (fit) J.K·mol$^{-1}$ | A (fit) J.K$^{-4}$mol$^{-1}$ | $\beta_3 - A\ln(T_{MFL})$ (fit) J.K$^{-4}$mol$^{-1}$ | $\beta_3$ (High $T$ saturation value) J.K$^{-4}$mol$^{-1}$ |
|---|---|---|---|---|---|
| 0 | 39.2 ± 1.0 | 22.18 ± 4.20 | (8.75 ± 0.23)× 10$^{-4}$ | -0.0019 ± 0.0002 | 2.18× 10$^{-5}$ |
| 1 | 37.5 ± 1.2 | 21.15 ± 3.59 | (7.79 ± 0.44)× 10$^{-4}$ | -0.0016 ± 0.0001 | 1.87× 10$^{-5}$ |
| 4 | 39.0 ± 1.0 | 20.52 ± 2.79 | (8.31 ± 0.42)× 10$^{-4}$ | -0.0018 ± 0.0001 | 2.61× 10$^{-5}$ |
| 7 | 45.1 ± 0.5 | 20.27 ± 3.33 | (6.38 ± 0.43)× 10$^{-4}$ | -0.0012 ± 0.0003 | 1.62× 10$^{-5}$ |
| 9 | 37.8 ± 1.1 | 22.12 ± 4.37 | (7.49 ± 0.36)× 10$^{-4}$ | -0.0015 ± 0.0002 | 2.62× 10$^{-5}$ |

## (b) Fitting parameters of one single expression over the entire $T$ range of $C(T)$ measurement

Furthermore we would like to mention that instead of doing fits to specific heat data in different $T$ regimes (as shown in Fig. 1(c) of main paper) we show in Fig. 4(b)-(i) below that the single expression $C(T) = \gamma T + \beta_3 T^3 + A T^3 \ln\left(\frac{T}{T_{MFL}}\right) + B T^{-2}$, fits over the entire $T$ regime for $C(T)$. The fitting parameters for whole range, single equation fitting are given below.

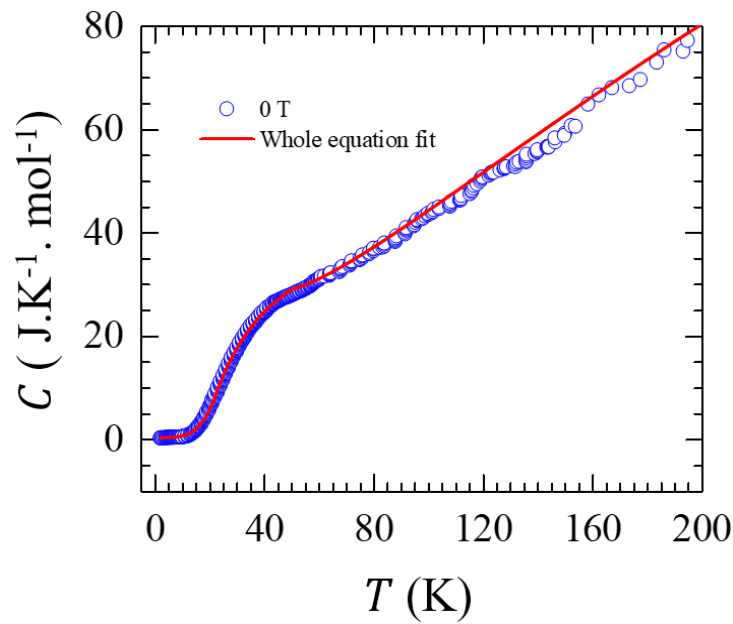

Fig.4(b) (i) Whole equation fit of $C(T)$ for the entire $T$ range

| Parameter | Value with Fitting done using the entire $T$ range |
|---|---|
| $\gamma$ | $41.2 \pm 0.5$ mJ. K$^{-2}$. mol$^{-1}$ |
| $A$ | $(8.61 \pm 0.18) \times 10^{-4}$ J. K$^{-4}$. mol$^{-1}$ |
| $B$ | $25.27 \pm 2.58$ J. K. mol$^{-1}$ |
| $\beta_3$ | $2.18 \times 10^{-5}$ J. K$^{-4}$. mol$^{-1}$ |

## (c) Estimating the in-gap value $\phi_s$ from specific heat and its comparison with that determined from $I - V$

At low $T$ the specific heat $C(T)$ of SmB$_6$ does not follow a purely electronic $\gamma T$ behavior, instead between 5-10 K, $C(T)$ shows a clear Schottky Anomaly feature of the form $BT^{-2}$. This $BT^{-2}$ behaviour can be obtained from an analysis of the specific heat of two-level system with a gap $\phi_s$, where the Schottky anomaly term of specific heat behaves as :

$$C_{Schottky} = Nk_B \left(\frac{\phi_s}{T}\right)^2 \frac{e^{\phi_s/T}}{\left(1 + e^{\phi_s/T}\right)^2}$$

where $\phi_s$ is the in-gap, $N$ is the no. of particles. One can readily note that in the limit of $\frac{\phi_s}{T} \ll 1$, one gets the form $C \sim BT^{-2}$ where $B \sim Nk_B \phi_s^2$, within the above model. Using the $B$ value determined from our Schottky fit of $C(T)$ (Fig.1(c) in the main MS) we estimate an in - gap value of $\phi_s$= 0.28 ± 0.12 meV. It must be mentioned that a more accurate estimate of the in-gap value $\phi_s$ from the Schottky anomaly feature in $C(T)$ needs to incorporate complex multilevel contributions to model the $C_{Schottky}$ [8], which is currently beyond the scope of the present work due to the complex band structure of SmB$_6$.

Therefore, the in-gap value of 0.28 meV of SmB$_6$ determined from the Schottky anomaly in $C(T)$ is within an order of magnitude agreement with the in-gap value estimated from $I$-$V$ (2.22 meV), This agreement is reasonable, given the experimental and analytical (model dependent) uncertainties.

# SECTION V:
## DC magnetic susceptibility $\chi(T)$ response of $s^{II}$ of $SmB_6$

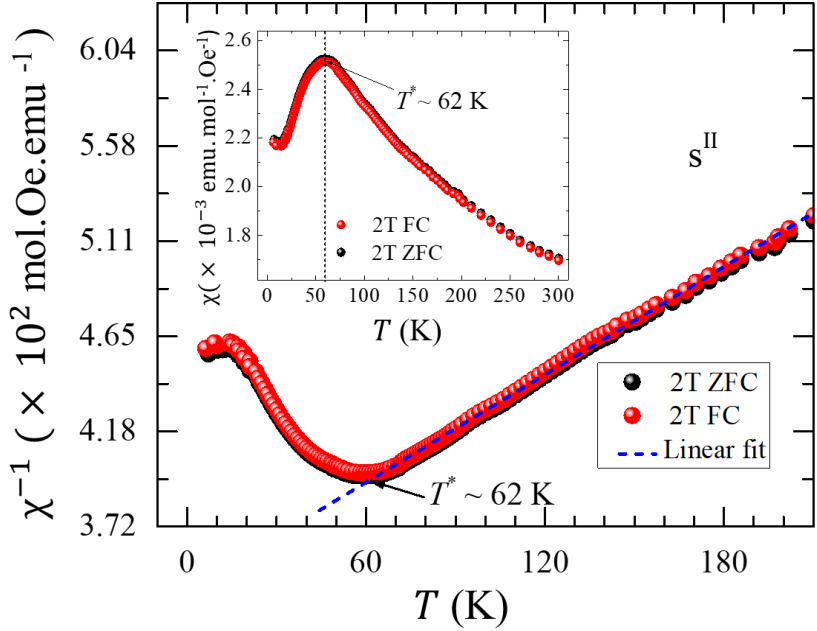

**Fig. 5.** $\chi^{-1}(T)$ response of $SmB_6$ $s^{II}$ at 2 T field (ZFC and FC) showing a linear Curie-Weiss fit for $T > T^*$. Inset shows the $\chi(T)$ response for 2 T ZFC and FC in $SmB_6$ ($s^{II}$).

Fig. 5 inset shows the DC magnetic susceptibility ($\chi(T)$) response of $SmB_6$ ($s^{II}$) for both zero-field cooled (ZFC) and field cooled (FC) conditions at 2 T magnetic field. ZFC and FC data show no bifurcation, indicating absence of irreversible or history dependent magnetism in $SmB_6$. We do not find any evidence of long-range magnetic correlations. Magnetic hysteresis is not observed at any $T$ in our crystals. The dome shaped feature around 62 K suggests the onset of hybridization by Kondo screening of magnetic moments, which is indicated as $T^*$. In the main panel of Fig. 5, we show the inverse susceptibility response i.e. $\chi^{-1}(T)$ for ZFC and FC data of $SmB_6$ $s^{II}$. At high $T$ (> 60 K), the data fits (blue dashed line) the Curie-Weiss law, i.e., $\chi(T) = \frac{C}{T-\theta}$, with $\theta = -386.2$ K and Curie constant $C = 1.14$ emu.K.mol$^{-1}$.Oe$^{-1}$. The nature of the magnetization data reported for $s^{II}$ is similar to that of $s^{I}$ reported in our earlier work [3], as well as excellently consistent with previously reported magnetization data in $SmB_6$ by other groups [9,10]. Presence of strong correlation effects above the bulk Kondo gap temperature can lead to the observation of a Curie Weiss like behaviour as in strongly correlated itinerant electron systems in the absence of any magnetic ordering [10-14]. The linear Curie-Weiss fit deviates significantly below $T^* \sim 62$ K where there is a hump. The regime below the hump in susceptibility has been associated with a regime with magnetic fluctuations [9].

# SECTION VI:

## (a) Negligible Joule heating effects in transport measurements

It has been ensured that there are negligible Joule heating effects on our transport measurements. The reasons are following:

1. Had Joule heating been a dominant factor in our transport measurement, there would have been significant thermal drifts in the measurement. Consequently, the values of $T_g$ and $T^*$ we obtain from transport measurements wouldn't have matched with those we determine from other measurements. We report a similarity of the values of $T_g$ and $T^*$ which have been identified from two completely different measurements viz., one based on electrical transport $\rho(T)$ (Fig.1(b) of main paper) which may have Joule heating effects and the other, a non-contact TCMI measurement (Fig.1(a) of main paper) where these effects are minimal. In fact, value of $T^* \sim 62$ K seen from DC SQUID susceptibility measurement shown in supplementary section V also matches very well with that obtained from electrical transport and TCMI measurement. Such close agreement between the $T$ location of different characteristic features in different measurements would not be possible had there been significant joule heating effects. This suggests that Joule heating effects are not a major contributing factor in our measurements.

2. Fig. 2(a) of main paper shows that at 5 K, a $I$ of 5 mA gives a <u>$V \sim 0.8$ mV</u>, which means the <u>$R \sim 0.16\ \Omega$</u> and the power dissipated is about <u>$P = 4\ \mu W$</u>, which is very small. While our sensors are not directly on the sample, but they are close to ~1.0 mm from the sample on a high conducting puck and here too we do not see any large $T$ fluctuations after waiting for some time after sending in current into the sample. In this measurement we observed a typical $\frac{<\Delta T>}{T} \leq 0.005$ where $<\Delta T>$ is the average $T$ fluctuations. The $\frac{<\Delta T>}{T} \leq 0.005$ was seen to be independent of $I$ variation up to 10 mA on the sample (10 mA is above the range of $I$ variation in our $I$ - $V$ measurements).

3. If Joule heating were to be a dominant factor in our transport measurements it would not have been possible to see the straight linear trend in $ln(I_s/T^2)$ vs. $1/T$ spanning a temperature from ~ 20 K to 5 K in the data of Fig. 2(d) in main paper, which have been obtained from transport measurements. There would have been a significant deviation from the linear trend in this figure especially at the higher $T$ (due to differing resistivity at different $T$) values. We do not observe anything like this.

4. A smooth fit to the $I$ - $V$ data with the Schottky equation for both samples of $SmB_6$ $s^I$ and $s^{II}$ (Fig. 2(c) main paper and SI section VI(b) below) with different IRR values showing similar $I$ - $V$ characteristics, indicates robustness of the measurements and negligible influence of any Joule heating effects.

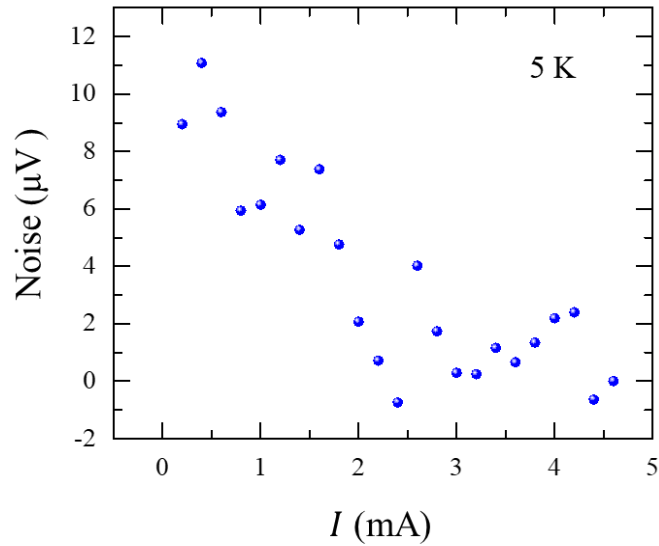

**Fig. 6(a). Noise in *I-V* at 5 K**

Figure 6(a) shows the fluctuations in $V$ at different $I$ at 5 K. The noise in $V$ vs $I$ above, show that for entire current range there is no significant changes in the fluctuations. If Joule heating were present at higher currents, the voltage noise fluctuations would have been higher. However, the observed low level in noise even with higher current indicates that Joule heating has no influence on our $I$ - $V$ data at 5 K.

### (b) DC drive study of $SmB_6$ sample $s^I$:

Fig. 6(b) shows the external DC drive studies on $SmB_6$ $s^I$ i.e. which we had used in our earlier work [3]. For this sample, $T_g = 40$ K and $T^* = 66$ K. From Fig. 6(b)-(i) and (ii) we can see, $SmB_6$ ($s^I$) shows non-linear $I$ - $V$ characteristics at $T < T_g$ which becomes a linear metal-like $I$ - $V$ characteristics at $T > T_g$ (Fig. 6(b)-(i) inset). The non-linearity feature seems to reduce as we approach closer to 40 K ~ $T_g$ (Fig. 6(b)-(ii)) and entirely vanishes as soon as we cross

$T_g$ (Fig. 6(b)-(i) inset). Fig. 6(b)-(iii) shows $I$ - $V$ at 8 K with the red dashed line showing the fit of the Schottky equation $I = I_s[exp\left(\frac{eV}{k_BT}\right) - 1]$ for 8 K. In Fig. 6(b)-(iv) the $ln(\frac{I_s}{T^2})$ vs $T^{-1}$ plot shows a linear fit (red dashed line), where from the slope we calculate $\varphi_s = 4.20 \pm 0.10$ meV.

These results conclude that $s^I$ behaviour with DC drive is exactly consistent with that of $s^{II}$.

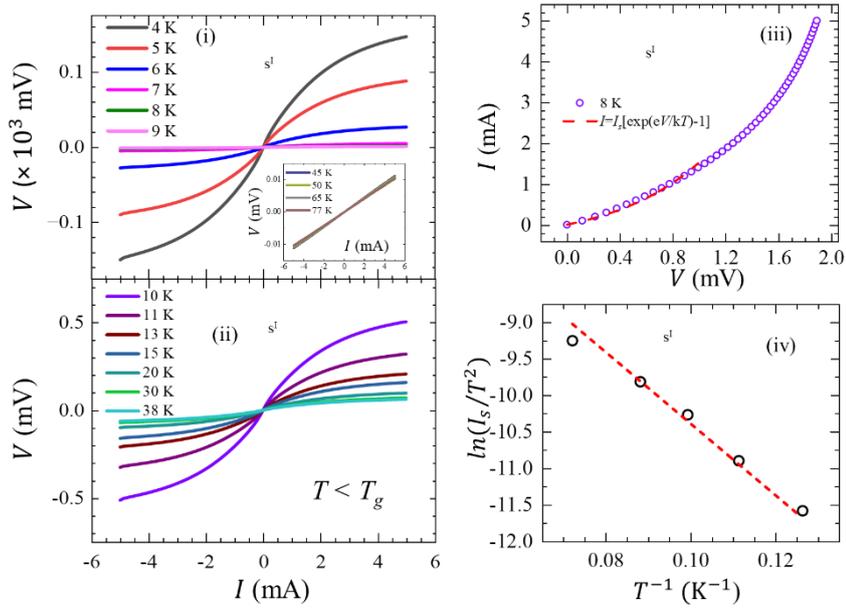

**Fig. 6(b). DC drive study on SmB$_6$ s$^I$:** (i) and (ii) shows $I$ - $V$ at $T < T_g$. Inset in (i) shows $I$ - $V$ at $T > T_g$. (iii) Schottky fit for s$^I$ at 8 K. (iv) $\varphi_s$ gap calculation from $ln(\frac{I_s}{T^2})$ vs $T^{-1}$ plot.

# SECTION VII:

## (a) Residual analysis to determine $\nu_0$ from $V'(\nu)$ plots for SmB$_6$ s$^{II}$

While the location of $\nu_0$ is clearly discernible from the prominent cusp in $|V''(\nu)|$ (see Fig. 3(a) of the main paper), its location in $V'(\nu)$ (Fig. 3(b) in the main paper) appears in a frequency regime where there is a crossover from bulk response with $V'(\nu) \propto \nu^2$ behavior to a regime with dominant surface conduction response showing $V'(\nu) \propto \nu^{0.4-0.6}$.

In Fig. 7(a) (i) and (ii) below, we show a residual analysis i.e., $|V'_{data} - V'_{fit}|$ vs $\nu$ at 30 K and 45 K respectively to show the goodness of the fit $V'(\nu) \propto \nu^2$ (in bulk dominated regime) and $V'(\nu) \propto \nu^{0.4}$ or $\nu^{0.6}$ (in surface dominated regime). Here $V_{fit}$ is the voltage value calculated using the fitting formula $V'(\nu) \propto \nu^2$ or $V'(\nu) \propto \nu^{0.6}$ (for 30 K) and $V'(\nu) \propto \nu^{0.4}$ (for 45 K). $V'_{data}$ is taken from the actual $V'(\nu)$ data of 30 K ad 45K from Fig. 3(b) of the main paper. In the residue plots, the deviation from zero marks $\nu_0$ i.e. the crossover frequency in the behavior of $V'(\nu)$. Clearly we can identify that the crossover in $V'(\nu)$ response from $\nu^2$ to $\nu^{0.6}$ (for 30 K) and $\nu^{0.4}$ (for 45 K) behavior occurs in the vicinity of $\nu_0$ (brown dotted lines in Fig. 7(a) below) where there is a sharp cusp in $|V''(\nu)|$ (Fig. 3(a) of main paper). $V'(\nu) \propto \nu^2$ fits well in low $\nu < \nu_0$ regime (yellow regime where residue of $\nu^2$ fit $\to 0$) and $V'(\nu) \propto \nu^{0.4-0.6}$ fits well in higher $\nu > \nu_0$ regime (blue regime where residue of $\nu^{p=0.4\ or\ 0.6}$ fit $\to 0$). This identification technique of $\nu_0$ is consistently followed for all the $V'(\nu)$ and $|V''(\nu)|$ plots at different $T$ for both s$^I$ and s$^{II}$ as shown in the main paper as well as supplementary material.

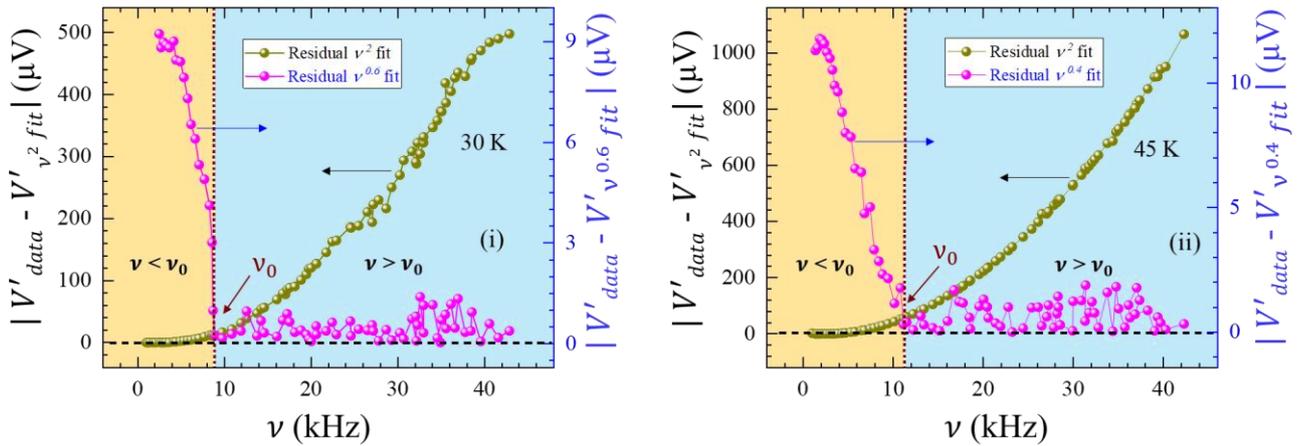

Fig. 7(a). Residual analysis of $V'(\nu)$ in SmB$_6$ s$^{II}$ for (i) $\nu^2$ (dark yellow data, bottom-X, left-Y in black) and $\nu^{0.6}$ (magenta data, top-X, right-Y in blue) fit at 30 K (ii) $\nu^2$ (dark yellow data, bottom-X, left-Y in black) and $\nu^{0.4}$ (magenta data, top-X, right-Y axis in blue) fit at 45 K. The maroon dotted line marks the location of $\nu_0$.

## (b) $|V''(\nu)|$ plots at different $T$ for SmB$_6$ s$^{II}$

Fig. 7(b) below presents the $|V''(\nu)|$ plots at different $T$ for SmB$_6$ s$^{II}$. While the location of $\nu_0$ can be clearly identified from the sharp cusp-like feature in the plots at higher $T$ ($T > T_g$) i.e. from 25K and above (see Fig. 7(b)- (vii) and (viii) below), as we decrease the $T$, only a broad minima in $|V''|$ is present at $\nu_0$ (see Fig. 7(b)- (iv), (v) and (vi) below). Finally, due to the dominant screening of the highly conducting surface layer over most of the bulk from the AC electromagnetic field, the feature of the bulk to surface crossover point $\nu_0$ is completely gone at even lower $T$ (Fig. 7(b)- (i), (ii) and (iii) below).

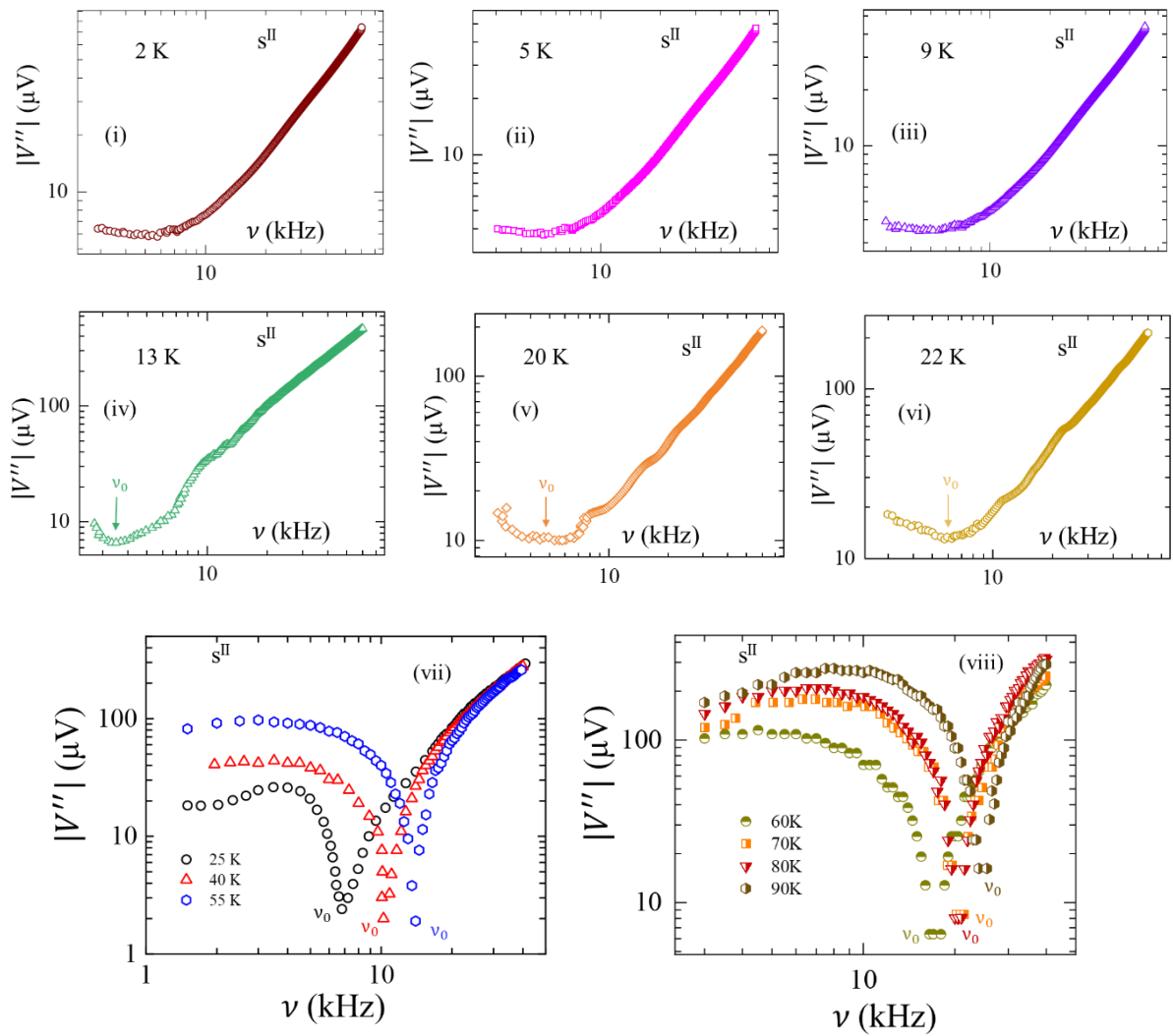

**Fig. 7(b).** $|V''(\nu)|$ plots in SmB$_6$ s$^{II}$ at different $T$ from 2 K up to 90 K. $\nu_0$ has been identified from broad minima or sharp cusp feature from 13 K and above.

### (c) $V'(\nu)$ plots at different $T$ for SmB$_6$ s$^{II}$

Fig. 7(c) below shows the $V'(\nu)$ plots at different $T$ for SmB$_6$ s$^{II}$. For all $T < T_g$, where the surface conduction dominates over bulk, a consistent $V'(\nu) \propto \nu^{0.6}$ dependence is observed over the entire frequency range from 2 K up to 22 K (see Fig. 7(c)- (i)). For all $T > T_g$, there is a crossover from bulk to surface-dominant behavior at $\nu_0$, where the behavior changes from $V'(\nu) \propto \nu^2$ (bulk) at lower $\nu$ to $V'(\nu) \propto \nu^{0.4-0.6}$ (surface) for higher $\nu$ (see Fig. 7(c)-(ii),(iii)). The location of $\nu_0$ at different $T$, presented here in $V'(\nu)$, is consistent with the $\nu_0$ values marked in $|V''(\nu)|$ plots at the same respective $T$ plots earlier (section 7(b) of supplementary).

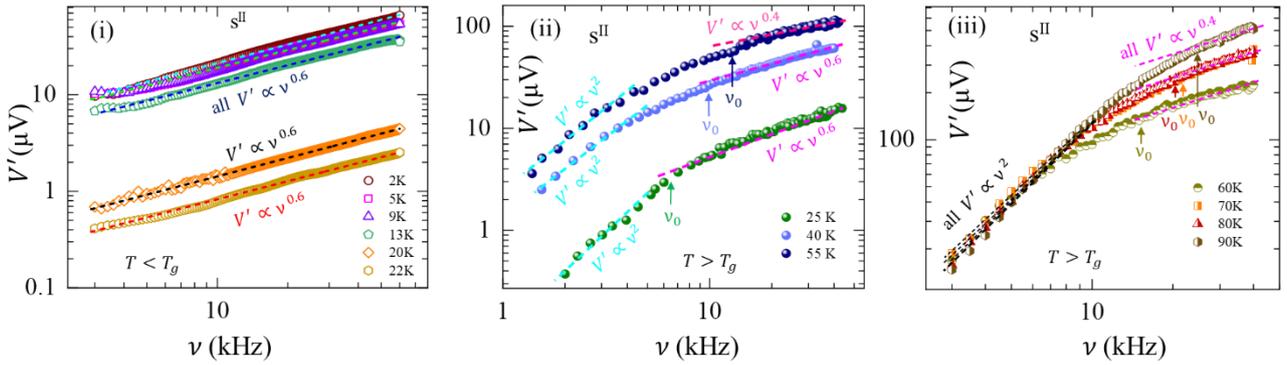

**Fig. 7(c).** $V'(\nu)$ **plots in SmB$_6$ s$^{II}$ at different $T$ from 2K up to 90 K. $\nu_0$ has been identified at all $T > T_g$.**

### (d) $|V''(\nu)|$ and $V'(\nu)$ plots at different $T$ for SmB$_6$ s$^{I}$

In Fig. 7(d)- (i) and (ii) below, we show the $|V''|$ and $V'(\nu)$ signal respectively of SmB$_6$ s$^{I}$ using the TCMI technique at two different temperatures i.e. 45 K (data in black circles) and 57 K (data in red triangles). Note that both the $T$ scales here are in the regime-II i.e. between $T^*$ (66 K) and $T_g$ (40 K) of SmB$_6$ s$^{I}$. The $\nu_0$ is marked at the sharp cusp-like feature in $|V''|$ at the values of 3.5 kHz and 5 kHz for $T = 45$ K (marked by black short-dashed line) and $T = 57$ K (marked by red short-dashed line) respectively. Fig. 7(b) - (ii) depicts that at $\nu < \nu_0$, both the temperatures show a $V'(\nu) \propto \nu^2$ bulk dominant behaviour (dark yellow dashed line fit) whereas at $\nu > \nu_0$, it shows $V'(\nu) \propto \nu^{0.4}$ (magenta dashed line fit at 45 K) or $\nu^{0.3}$ (cyan dashed line fit at 57 K). Therefore, our results for $s^{II}$ are exactly consistent with those observed and reported for $s^{I}$ of $SmB_6$ as shown here as well as in our earlier work [3].

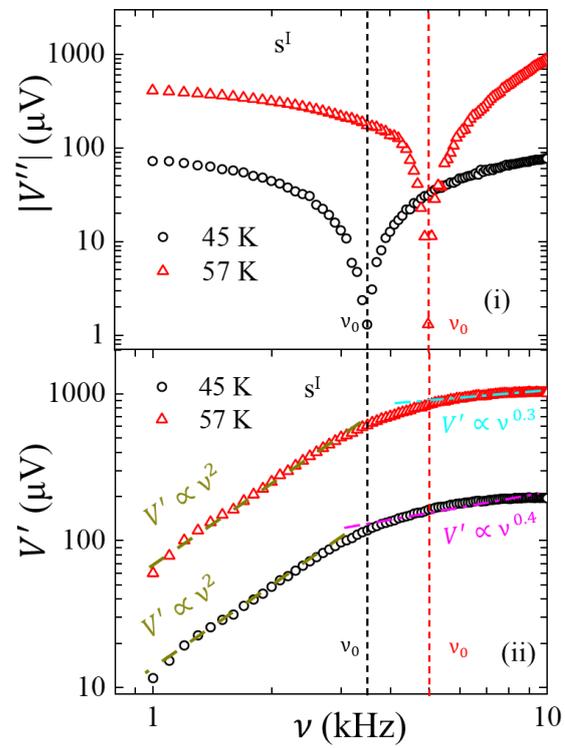

**Fig. 7(d). (i) $|V''|(\nu)$ (ii) $V'(\nu)$ plot in $SmB_6$ $s^{I}$ at 45 K and 57 K. $\nu_0$ has been identified by black and red dashed lines for 45 K and 57 K respectively.**

# SECTION VIII:
# Study of dissipation behavior from out-of-phase response of SmB$_6$ s$^I$ and s$^{II}$:

Fig. 8(i) and (ii) show the out of phase signal $V''(\nu)$ and the absolute value of that signal i.e. $|V''(\nu)|$ respectively for SmB$_6$ s$^{II}$ at three different temperatures i.e. 25 K (black circle), 40 K (red triangle) and 55 K (blue square) from regime (ii) i.e. $T_g < T < T^*$. In Fig. 8(i), we can observe the $V''(\nu)$ has a sign change below and above a certain characteristic frequency scale which is also $T$ dependent (zero crossing point marked by black, red and blue arrow for 25 K, 40 K and 55 K respectively). Since the out of phase AC signal from two-coil implies dissipation behaviour and dissipation can not be negative, we have represented it as an absolute $|V''(\nu)|$ plot in Fig. 8(ii) for s$^{II}$ with the characteristic frequency scale marked as $\nu_0$ for the three $T$ values. Similar kind of $|V''(\nu)|$ behavior for two other temperature scales: 30 K and 45 K has been shown in Fig. 3(a) of main paper. The zero crossing of $V''(\nu)$ (black dashed line in Fig. 8(i)) is originated from the sign reversal of the phase of the AC pick-up response signifying two different kinds of behaviour of the sample below and above $\nu_0$. This has been elaborated in detail in the description of Fig. 4 of main paper. We also present the $V''(\nu)$ and $|V''(\nu)|$ response of s$^I$ in Fig. 8(iii) and (iv) respectively showing a similar sign reversal of out of phase signal at a certain $\nu_0$ value which confirms our results to be consistent for both the samples.

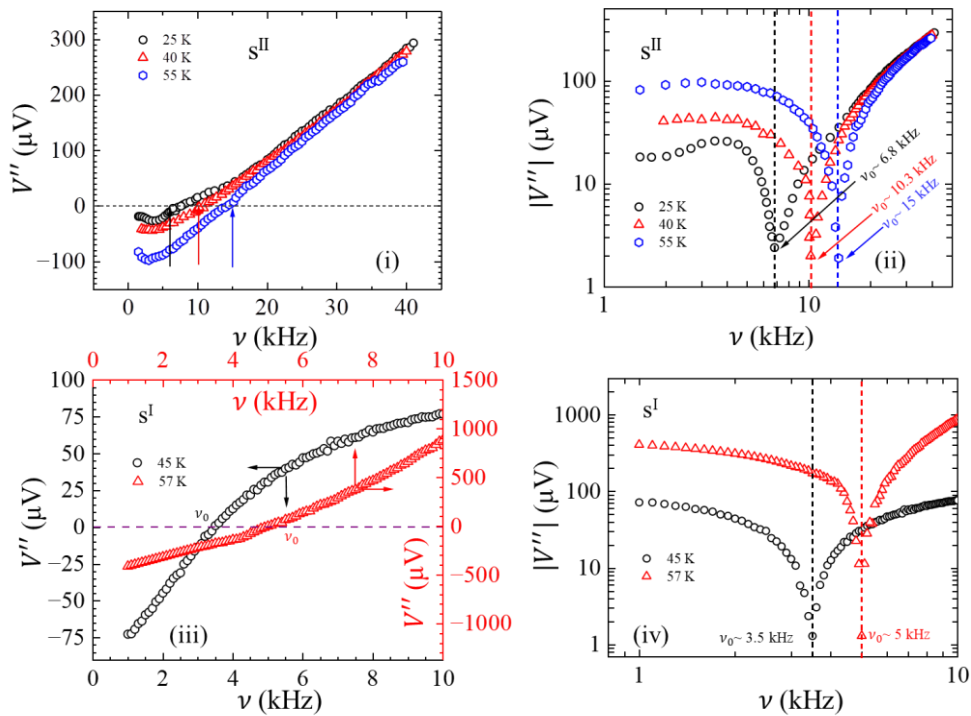

**Fig. 8. Dissipation study in SmB$_6$ s$^I$ and s$^{II}$:** (i) $V''(\nu)$, (ii) $|V''(\nu)|$ for s$^{II}$ at 25 K (black), 40 K (red), and 55 K (blue). (iii) $V''(\nu)$, (iv) $|V''(\nu)|$ for s$^I$ at 45 K (black) and 57 K (red).

In Fig. 8(iii), the $V''(\nu)$ is shown for two different temperatures from regime (ii) of s$^I$ i.e. 45 K (black circle, left-Y and bottom-X axes) and 57 K (red triangle, right-Y and top-X axes). wine dashed line marks the zero-crossing line of $V''(\nu)$ for both the temperatures. The exact value of $\nu_0$ has been marked as 3.5 kHz (black dashed line and arrow) and 5 kHz (red dashed line) from the sharp cusp-like feature in $|V''(\nu)|$ at 45 K and 57 K respectively in s$^I$ (see Fig. 8(iv)).

## SECTION IX:
## Frequency varied phase (ϕ) response at different $T$ of $s^{II}$ by TCMI technique:

Figure 9 shows the phase ($\phi = \tan^{-1}\left(\frac{V''}{V'}\right)$) between the in-phase $V'$ and out of phase $V''$ components of the induced voltage as a function of excitation frequency ($\nu$) at different $T$ measured by two-coil technique on SmB$_6$ $s^{II}$. It shows that for all $T$ in regime II ($T_g < T < T^*$) i.e. 23 K to 62 K, the phase changes sign from negative to positive at a particular $\nu_0$ (marked by black arrows in Fig. 9) implying the arrival of new inductive components related to surface layer conductance at $\nu > \nu_0$ inside the system. $\nu_0$ shifts to higher values when $T$ increases in this regime. But regime-I ($T > T^*$) i.e. 294 K being entirely bulk dominant at such high $T$ and regime-III ($T < T_g$) i.e. 17 K having dominant surface layer conductivity response, do not show any zero-crossing point or $\nu_0$ in the phase modulation. This is expected since in these regimes only a single type of behaviour dominates instead of an admixture phase being modulated by frequency variation (high $\nu$: surface, low $\nu$: bulk) which is only present for regime-II. This can be more clearly understood by the equivalent electrical circuit modelling by Nyquist plots of the $s^{II}$ (refer to Fig. 4 main manuscript).

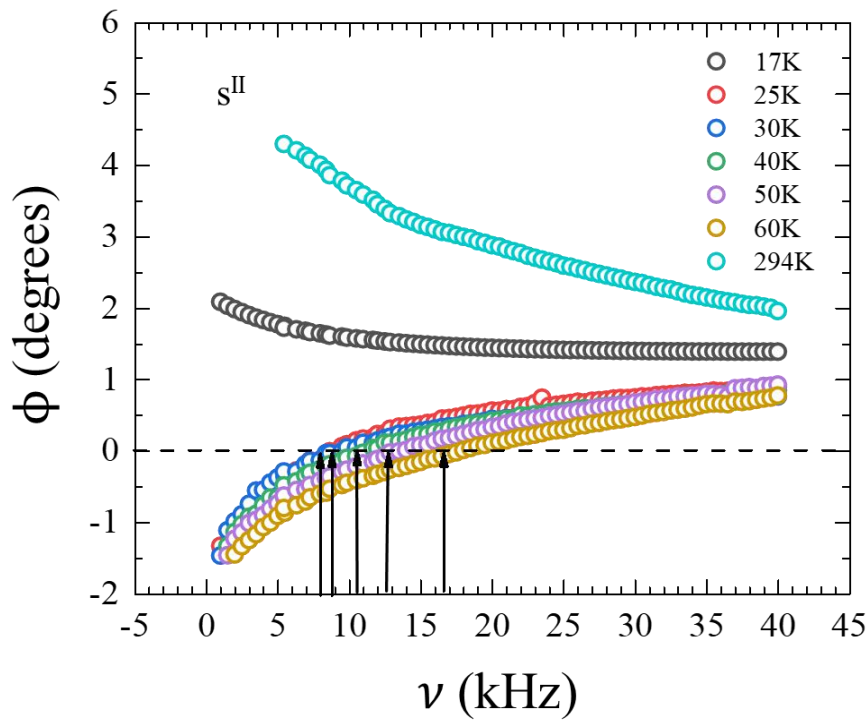

**Fig. 9.** Frequency varied phase (ϕ) study in SmB$_6$ $s^{II}$ at different $T$ regimes (regime-i: 294K, regime-ii: 25K, 30K, 40K, 50K, 60K and regime-iii: 17K)

# SECTION X:
# Equivalent Nyquist plot analysis of TCMI response at different $T$ for $s^{II}$

The $V'(\nu)$ and $V''(\nu)$ response from $s^{II}$ SmB$_6$ using TCMI technique at different $T$ has been modelled and matched with a simulated electrical circuit network by Nyquist analysis method. Nyquist plots used for this analysis are in the form of $V''(\nu)$ vs $V'(\nu)$ (left-Y, bottom-X axis) plot matched with simulated simulated equivalent circuit impedance components $-Z''(\nu)$ vs $Z'(\nu)$ plot (right-Y, top-X axis). In Fig. 10(a) below, the Nyquist plots for all $T < T_g$ (= 23 K), i.e. from 22 K down to 2 K (inset of panel i) in Fig. 10(a), also see Fig. 4(a) inset in main paper) is presented. It is observed that the data for $T < T_g$ is modelled by a single type of equivalent $R_1$- $R_2$-CPE- L model for the entire $\nu$ range. As we increase the $T$ above $T_g$, there is a crossover from one type of behavior to another, while going from $\nu < \nu_0$ to $\nu > \nu_0$ as the curves for both regimes are not analytically continuous across $\nu_0$ (where $V'' \to 0$). In the range of temperatures from 25 K to 90 K, for the bulk conductivity dominated regime, viz, $\nu < \nu_0$, we find a conventional R - C network doesn't model the Nyquist curve (Nyquist at 294 K in Fig. 4(a) of main paper is modelled by conventional R - C network), rather the curves are modelled well with two resistors: $R_1$, $R_2$ and a constant phase element (CPE) model (see panel no. i) in each of the Figs. 10. (b) – (k) shown below and also Fig. 4(b) in main paper). The impedance of CPE is, $Z_{CPE} = \frac{1}{Q_0(i\nu)^n}$ [15]. The modified $Z_{CPE}$ form is ascribed to non-ideal relaxation processes in a material, where $Q_0$ is the CPE capacitance and $n$ is the measure of degree of non-ideality from $\frac{1}{i2\pi\nu C}$ behaviour. At 25K – 90 K, in the surface conduction dominated regime i.e. $\nu > \nu_0$ (see panel no. ii) in Figs. 10. (b) – (k) shown below and also Fig. 4(c) in main paper), to model the Nyquist curve, a new reactive component: an inductor (L) has to be introduced along with CPE to simulate the plot nature. Note that the same component L is consistently present in all the Nyquist plots below $T_g$ and down to very low $T$ i.e. 2 K throughout the entire $\nu$ range as a signature of surface dominated conductivity response. The evolution of these components with $T$ (see Fig. 4(d), (e) in main paper) explains the physical significance of these for our sample as mentioned below in details:

**$R_1, R_2$ :** In our equivalent circuit model, $R_1$ represents the bulk resistance (hence rapidly rises at $T_g$), while $R_2$ represents the resistance associated with charged heavy carriers released on the surface due to broken Kondo state. Because of strong correlations and localization effects experienced by these charge carriers, the resistance $R_2$ increases with reducing $T$ below $T_g$. At higher $T$ (above $T_g$), $R_2$ represents normal metallic conductivity (See Fig. 4(d) in main paper).

**C (at high $T$) and CPE (at low $T$)**: As temperature decreases into the Kondo insulating regime, the bulk resistivity increases and a non-ideal capacitive behaviour emerges, which we model with a Constant Phase Element (CPE). With lowering $T$ (especially below $T^*$) the strengthening of strong correlation effects in $SmB_6$ necessitates the inclusion of CPE which represents a distributed capacitance (or a continuum of relaxation times). This non-ideal capacitive response arises from strong correlation effects causing development of a spectrum of time constants in the system (e.g., a distribution of trapped states or spatial variations in conduction due to defects). Such a term is often used for modelling impedance spectroscopy data (see added references in main paper). Below $T^*$, $Q_0$ (where $Z_{CPE} = \frac{1}{Q_0(iv)^n}$) shows a peak and below $T_k^s$, it saturates to a small but finite value (see Fig. 4(e) of main paper), where the topological surface conducting state develops in the backdrop of the broken Kondo state on the surface. At high $T$ as $n = 1$, $Q_0 \equiv C$.

**L**: L represents the reactive response of the high-mobility light surface quasiparticles. Physically, this inductive component (L) is related to the kinetic inductance of the topological surface charge carrier contribution to conductivity. This represents the screening behavior generated by the surface conduction region in response to the time varying magnetic field imposed on the sample in our two-coil measurement setup. In the newly added Fig. 4(e) in the revised main manuscript, below $T_g$, as the surface state contribution to conduction increases, the L also increases. Note below $T_k^s$, in the topological state, both L and $R_2$ contribute, though the increase in $R_2(T)$ with reducing $T$ is much slower. The simultaneous presence of L and $R_2$ at low $T$ is a characteristic feature of our two fluid state.

**Table-II: Nyquist parameters for equivalent circuit components at different $T$**

| $T$ | $R_1$ | $R_2$ | C | $Q_0$(CPE) | $n$(CPE) | L |
|---|---|---|---|---|---|---|
| 294 K | $1.000 \times 10^{-5}$ Ω | 40 Ω | $4 \times 10^{-6}$ F | × | × | × |
| 90 K, $\nu < \nu_0$ | $1.254 \times 10^{-5}$ Ω | 37 Ω | × | $3.0 \times 10^{-9}$ F | 1.10 | × |
| 90 K, $\nu > \nu_0$ | | | | $1.6 \times 10^{-9}$ F | | $1.8 \times 10^{-9}$ H |
| 80 K, $\nu < \nu_0$ | $1.254 \times 10^{-5}$ Ω | 35 Ω | × | $3.2 \times 10^{-9}$ F | 1.10 | × |
| 80 K, $\nu > \nu_0$ | | | | $1.8 \times 10^{-9}$ F | | $2.0 \times 10^{-9}$ H |
| 70 K, $\nu < \nu_0$ | $1.254 \times 10^{-5}$ Ω | 32 Ω | × | $3.8 \times 10^{-9}$ F | 1.10 | × |
| 70 K, $\nu > \nu_0$ | | | | $2.1 \times 10^{-9}$ F | | $2.5 \times 10^{-9}$ H |
| 60 K, $\nu < \nu_0$ | $1.254 \times 10^{-5}$ Ω | 30 Ω | × | $3.9 \times 10^{-9}$ F | 1.10 | × |
| 60 K, $\nu > \nu_0$ | | | | $3.0 \times 10^{-9}$ F | | $5.0 \times 10^{-9}$ H |
| 55 K, $\nu < \nu_0$ | $1.254 \times 10^{-5}$ Ω | 29 Ω | × | $4.3 \times 10^{-9}$ F | 1.12 | × |
| 55 K, $\nu > \nu_0$ | | | | $3.4 \times 10^{-9}$ F | | $6.0 \times 10^{-9}$ H |
| 50 K, $\nu < \nu_0$ | $1.271 \times 10^{-5}$ Ω | 29 Ω | × | $4.5 \times 10^{-9}$ F | 1.12 | × |
| 50 K, $\nu > \nu_0$ | | | | $3.5 \times 10^{-9}$ F | | $8.0 \times 10^{-9}$ H |
| 45 K, $\nu < \nu_0$ | $1.277 \times 10^{-5}$ Ω | 28 Ω | × | $4.8 \times 10^{-9}$ F | 1.14 | × |
| 45 K, $\nu > \nu_0$ | | | | $2.9 \times 10^{-9}$ F | | $9.0 \times 10^{-9}$ H |
| 40 K, $\nu < \nu_0$ | $1.285 \times 10^{-5}$ Ω | 26 Ω | × | $5.0 \times 10^{-9}$ F | 1.15 | × |
| 40 K, $\nu > \nu_0$ | | | | $2.5 \times 10^{-9}$ F | | $1.0 \times 10^{-8}$ H |
| 35 K, $\nu < \nu_0$ | $1.287 \times 10^{-5}$ Ω | 25 Ω | × | $5.2 \times 10^{-9}$ F | 1.15 | × |
| 35 K, $\nu > \nu_0$ | | | | $2.4 \times 10^{-9}$ F | | $1.0 \times 10^{-8}$ H |
| 30 K, $\nu < \nu_0$ | $1.284 \times 10^{-5}$ Ω | 25 Ω | × | $5.3 \times 10^{-9}$ F | 1.16 | × |
| 30 K, $\nu > \nu_0$ | | | | $2.0 \times 10^{-9}$ F | | $4.0 \times 10^{-8}$ H |
| 25 K, $\nu < \nu_0$ | $1.540 \times 10^{-5}$ Ω | 22 Ω | × | $5.5 \times 10^{-9}$ F | 1.16 | × |
| 25 K, $\nu > \nu_0$ | | | | $2.0 \times 10^{-9}$ F | | $8.0 \times 10^{-8}$ H |
| 22 K | $2.260 \times 10^{-5}$ Ω | 40 Ω | × | $9.8 \times 10^{-10}$ F | 1.15 | $9.1 \times 10^{-8}$ H |
| 20 K | $3.840 \times 10^{-5}$ Ω | 58 Ω | × | $9.0 \times 10^{-10}$ F | 1.15 | $9.6 \times 10^{-8}$ H |
| 17 K | $1.371 \times 10^{-4}$ Ω | 80 Ω | × | $8.0 \times 10^{-10}$ F | 1.16 | $9.0 \times 10^{-8}$ H |
| 13 K | $6.842 \times 10^{-4}$ Ω | 84 Ω | × | $5.0 \times 10^{-10}$ F | 1.15 | $3.0 \times 10^{-7}$ H |
| 9 K | 0.012 Ω | 90 Ω | × | $4.5 \times 10^{-10}$ F | 1.15 | $5.0 \times 10^{-7}$ H |
| 5 K | 0.285 Ω | 95 Ω | × | $3.0 \times 10^{-10}$ F | 1.15 | $6.0 \times 10^{-7}$ H |
| 2 K | 0.363 Ω | 100 Ω | × | $2.9 \times 10^{-10}$ F | 1.15 | $7.0 \times 10^{-7}$ H |

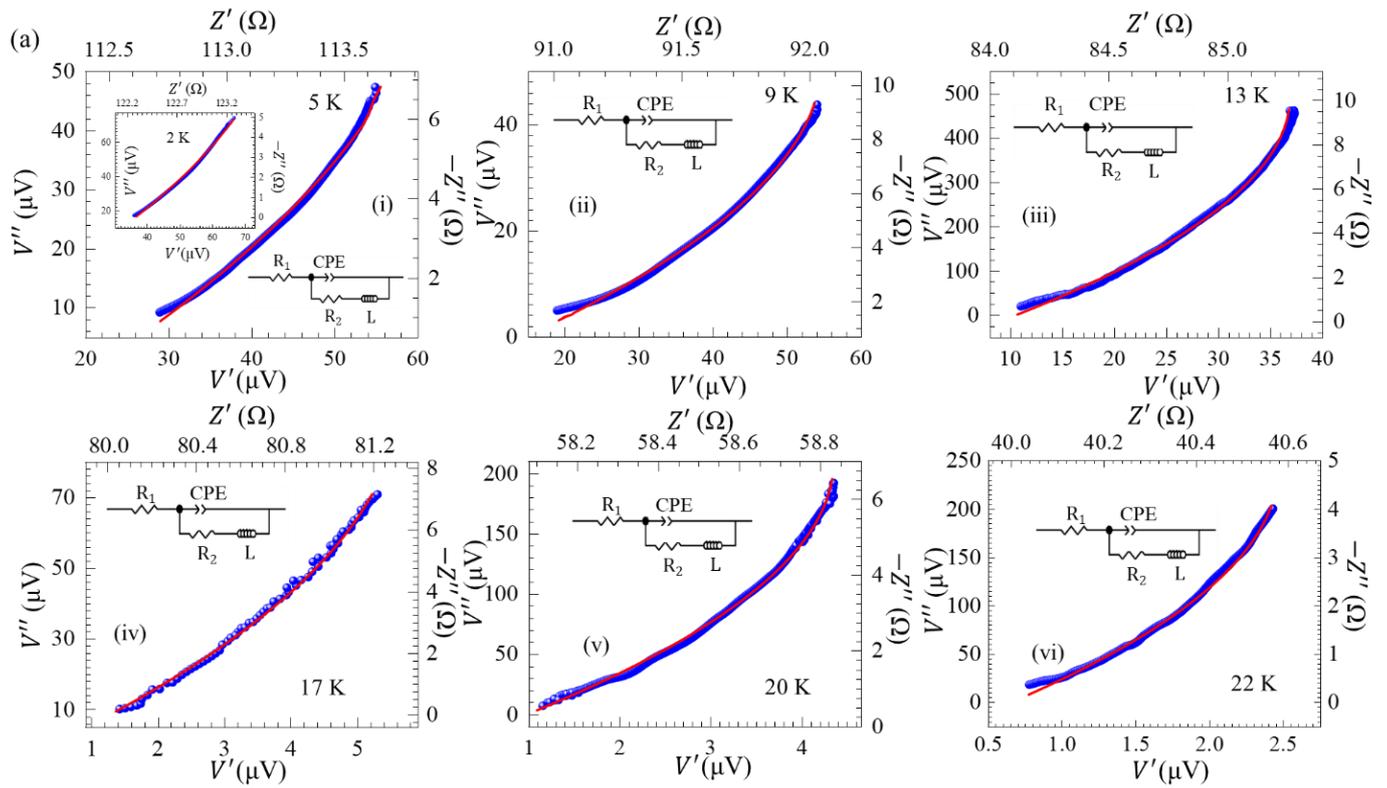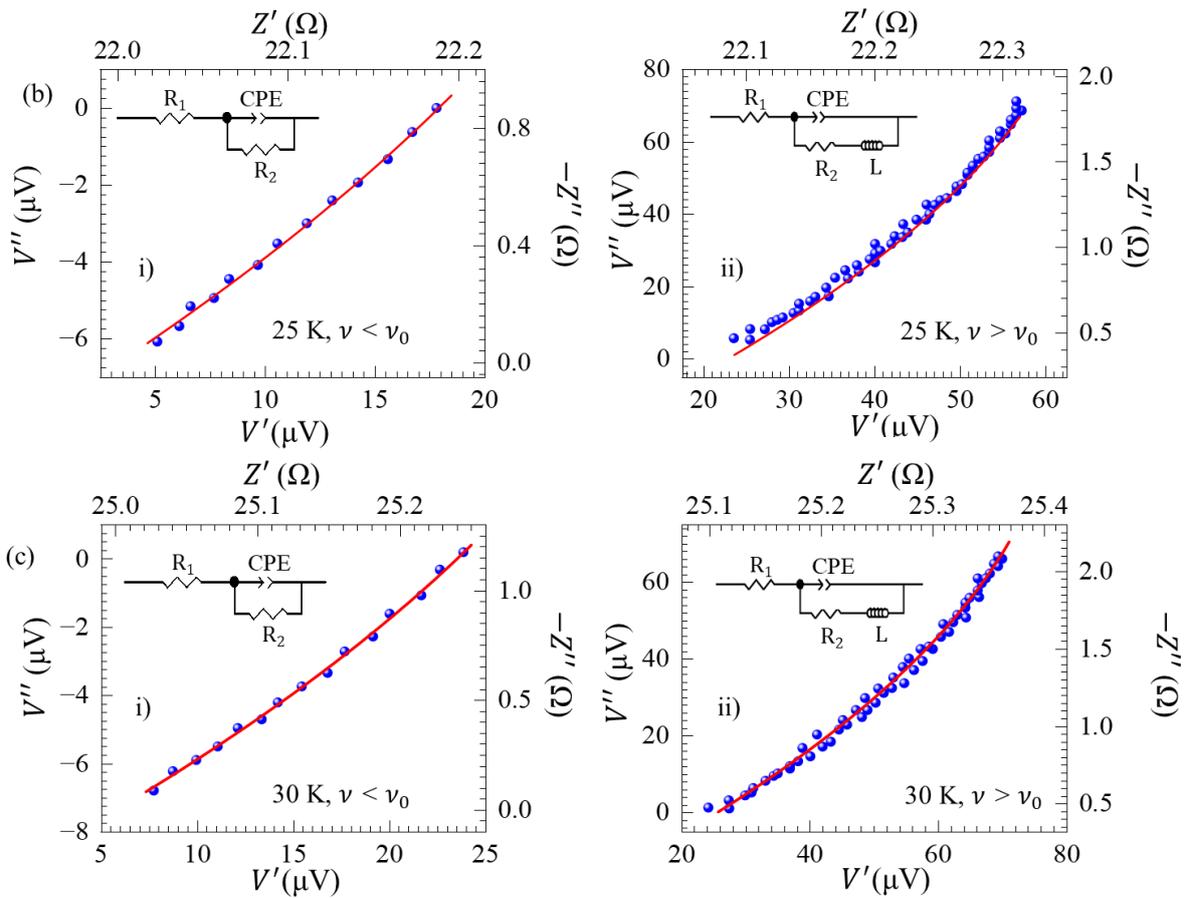

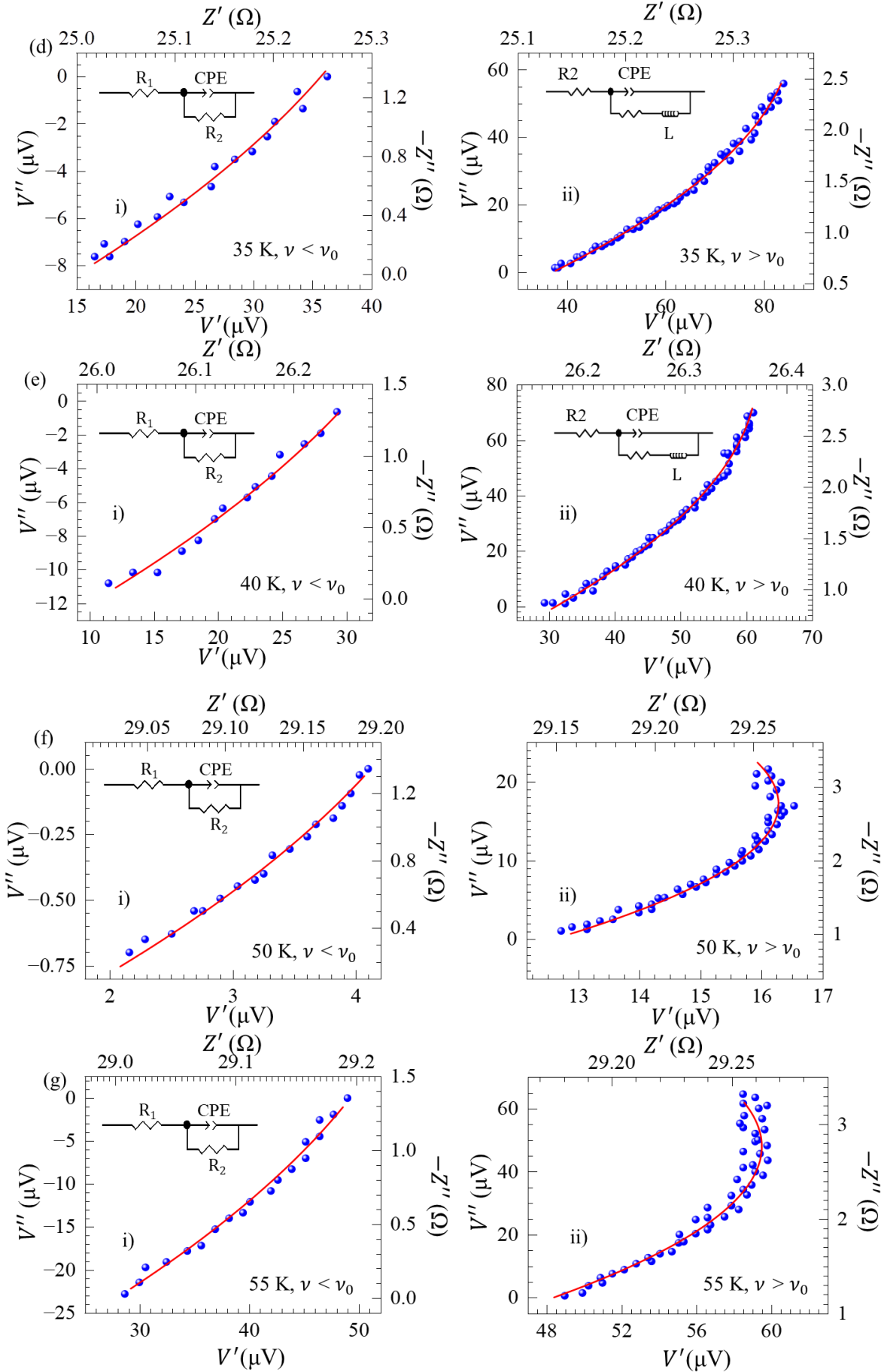

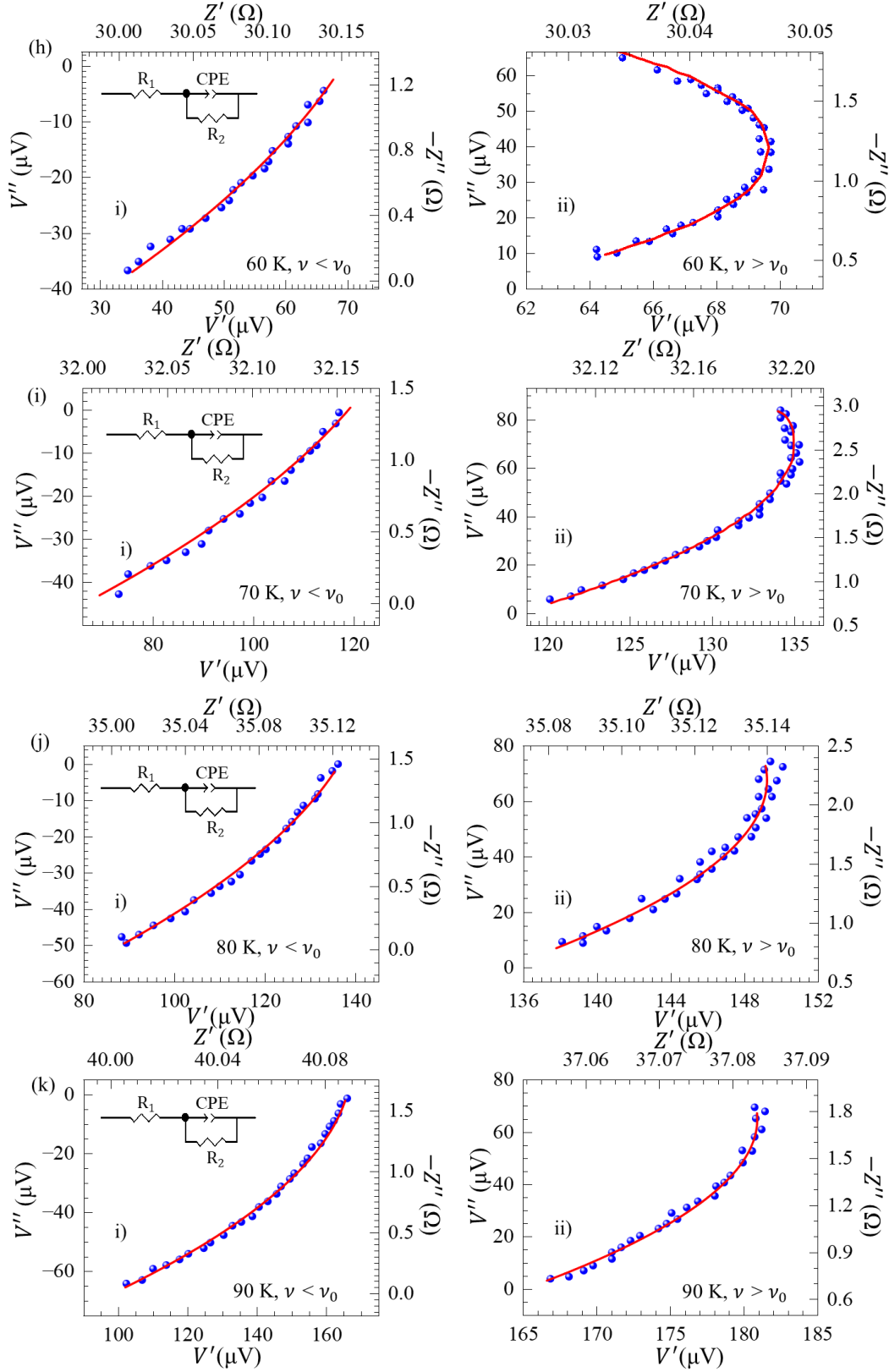

**Fig. 10.** Equivalent Nyquist analysis in SmB$_6$ s$^{II}$ at different $T$. **(a)** $T$ = 2 K to 22 K one single model for the entire $\nu$ regime. $T$ = 2 K Nyquist shown in panel i) inset. **(b) – (k)** from 25 K up to 90 K: two different models for the 2 different $\nu$ regimes (panel i. is for $\nu < \nu_0$ and ii. is for $\nu > \nu_0$ regime).

# REFERENCES

[1]	M. C. Hatnean, M. R. Lees, D. M. Paul, and G. Balakrishnan, Sci Rep **3**, 3071 (2013).
[2]	L. J. Berchmans, A. Visuvasam, S. Angappan, C. Subramanian, and A. K. Suri, Ionics **16**, 833 (2010).
[3]	S. Ghosh, S. Paul, A. Jash, Z. Fisk, and S. S. Banerjee, Physical Review B **108**, 205101 (2023).
[4]	A. Jash, K. Nath, T. R. Devidas, A. Bharathi, and S. S. Banerjee, Physical Review Applied **12**, 014056 (2019).
[5]	A. Jash, S. Ghosh, A. Bharathi, and S. S. Banerjee, Physical Review B **101**, 165119 (2020).
[6]	A. Jash, S. Ghosh, A. Bharathi, and S. S. Banerjee, Bulletin of Materials Science **45**, 17 (2022).
[7]	W. A. Phelan, S. M. Koohpayeh, P. Cottingham, J. W. Freeland, J. C. Leiner, C. L. Broholm, and T. M. McQueen, Physical Review X **4**, 031012 (2014).
[8]	P. S. Riseborough, Advances in Physics **49**, 257 (2000).
[9]	P. K. Biswas *et al.*, Physical Review B **89**, 161107 (2014).
[10]	W. Ruan, C. Ye, M. Guo, F. Chen, X. Chen, G.-M. Zhang, and Y. Wang, Physical Review Letters **112**, 136401 (2014).
[11]	V. Janiš, A. Klíč, J. Yan, and V. Pokorný, Physical Review B **102**, 205120 (2020).
[12]	K. Byczuk and D. Vollhardt, Physical Review B **65**, 134433 (2002).
[13]	J. M. Tomczak, K. Haule, and G. Kotliar, Proceedings of the National Academy of Sciences **109**, 3243 (2012).
[14]	I. A. Nekrasov, Z. V. Pchelkina, G. Keller, T. Pruschke, K. Held, A. Krimmel, D. Vollhardt, and V. I. Anisimov, Physical Review B **67**, 085111 (2003).
[15]	J. R. Macdonald, Annals of Biomedical Engineering **20**, 289 (1992).